\documentclass[manuscript]{acmart}

\usepackage{longtable}
\usepackage{fontawesome5}
\usepackage{tikz}
\usepackage{multirow}
\usepackage{tabularx}
\usepackage{tcolorbox}
\tcbuselibrary{skins, breakable}
\usepackage[]{mdframed}

\newcommand{\summaryBox}[2]{
	\begin{tcolorbox}[enhanced,
            attach boxed title to top left={yshift=-3mm,xshift=-0.5mm},
		colframe=black,
		leftrule=.1mm,toprule=.1mm,
		rightrule=.5mm, bottomrule=.5mm,
		colback=white,
		boxsep=1mm,
		top=3mm,
		arc=0mm,
		colbacktitle=white,
		coltitle=black,
		boxed title style={
			boxrule=0pt,
			colframe=white,
			coltitle=darkgray
		},
            title=\textbf{Implication~#1},
		width=\linewidth
		]
		{#2}
	\end{tcolorbox}
}

\AtBeginDocument{%
  }

\setcopyright{acmlicensed}
\copyrightyear{2025}
\acmYear{2025}
\acmDOI{XXXXXXX.XXXXXXX}

\acmISBN{978-1-4503-XXXX-X/18/06}

\acmISBN{978-1-4503-XXXX-X/18/06}

\begin{document}


\title[Usable Privacy in Virtual Worlds]{Usable Privacy in Virtual Worlds: Design Implications for Data Collection Awareness and Control Interfaces in Virtual Reality}


\author{Viktorija Paneva}
\email{viktorija.paneva@ifi.lmu.de}
\orcid{0000-0002-5152-3077}
\affiliation{%
  \institution{LMU Munich}
  \city{Munich}
  \country{Germany}}
\affiliation{%
  \institution{University of the Bundeswehr}
  \city{Munich}
  \country{Germany}}

\author{Verena Winterhalter}
\email{verena.winterhalter@unibw.de}
\orcid{0000-0003-0752-3480}
\affiliation{%
  \institution{LMU Munich}
  \city{Munich}
  \country{Germany}}
\affiliation{%
  \institution{University of the Bundeswehr}
  \city{Munich}
  \country{Germany}}

\author{Naga Sai Surya Vamsy Malladi}
\email{naga.malladi@uni-bayreuth.de}
\orcid{0009-0008-8957-8103}
\affiliation{%
    \institution{University of Bayreuth}
    \city{Bayreuth}
    \country{Germany}
    }

\author{Marvin Strauss}
\email{marvin.strauss@uni-due.de}
\orcid{0009-0007-1040-9175}
\affiliation{%
    \institution{Human-Computer Interaction Group, University of Duisburg-Essen}
    \city{Essen}
    \country{Germany}
    }

\author{Stefan Schneegass}
\orcid{0000-0002-0132-4934}
\email{stefan.schneegass@uni-due.de}
\affiliation{%
    \institution{Human-Computer Interaction Group, University of Duisburg-Essen}
    \city{Essen}
    \country{Germany}
    }

\author{Florian Alt}
\orcid{0000-0001-8354-2195}
\affiliation{%
  \institution{LMU Munich}
  \city{Munich}
  \country{Germany}}
\affiliation{%
  \institution{University of the Bundeswehr}
  \city{Munich}
  \country{Germany}}
\email{florian.alt@ifi.lmu.de}


\renewcommand{\shortauthors}{Paneva et al.}


\begin{abstract}
Extended reality (XR) devices have become ubiquitous. They are equipped with arrays of sensors, collecting extensive user and environmental data, allowing inferences about sensitive user information users may not realize they are sharing.
Current VR privacy notices largely replicate mechanisms from 2D interfaces, failing to leverage the unique affordances of virtual 3D environments. 
To address this, we conducted brainstorming and sketching sessions with novice game developers and designers, followed by privacy expert evaluations, to explore and refine privacy interfaces tailored for VR. 
Key challenges include balancing user engagement with privacy awareness, managing complex privacy information with user comprehension, and maintaining compliance and trust.
We identify design implications such as thoughtful gamification, explicit and purpose-tied consent mechanisms, and granular, modifiable privacy control options.
Our findings provide actionable guidance to researchers and practitioners for developing privacy-aware and user-friendly VR experiences. 
\end{abstract}


\begin{teaserfigure}
    \centering
\vspace{-4mm}
    \begin{tikzpicture}
    \draw (0, 0) node[inner sep=0] {
    \includegraphics[width=0.5\linewidth]{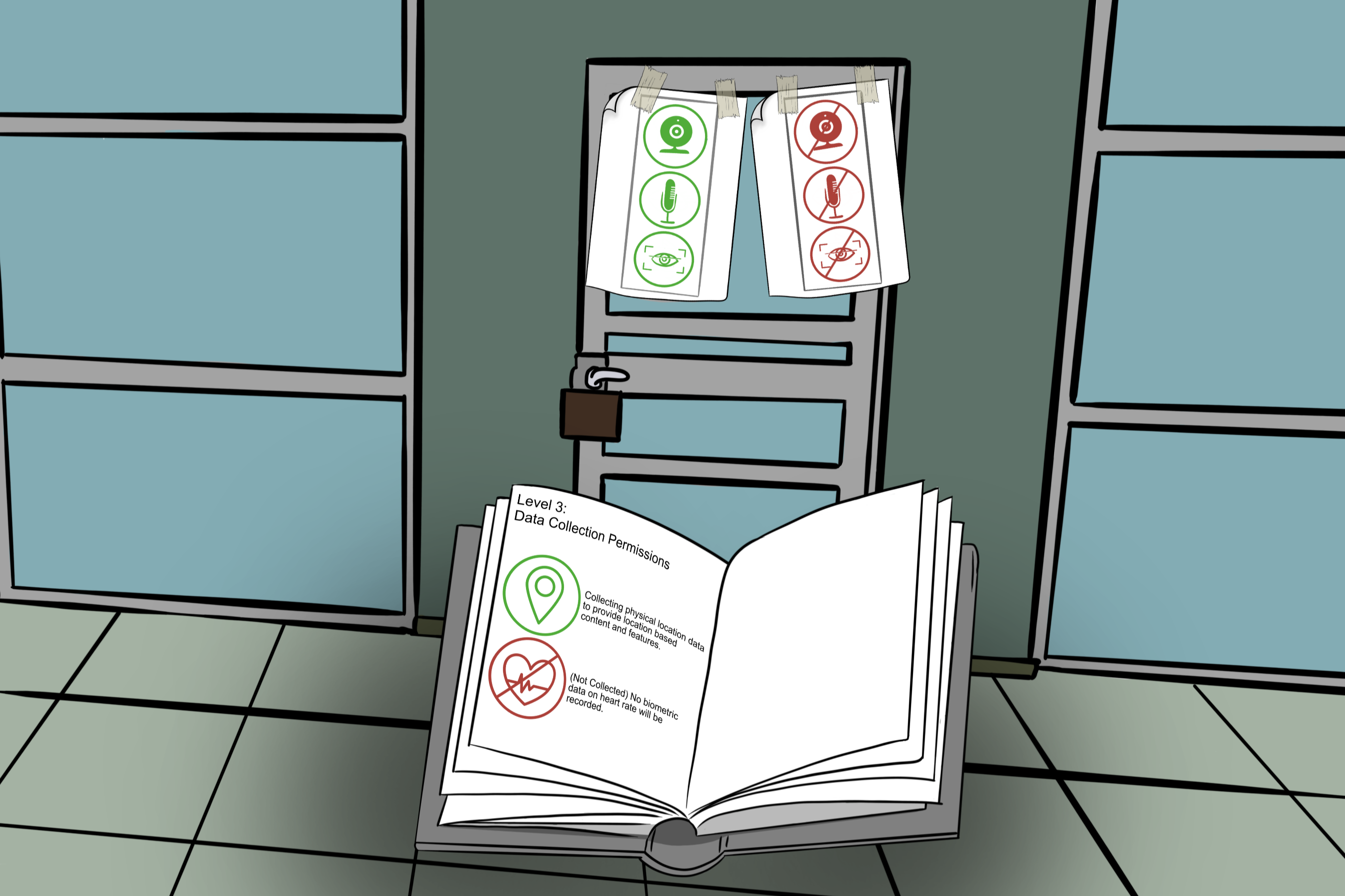}
    \includegraphics[width=0.5\linewidth]{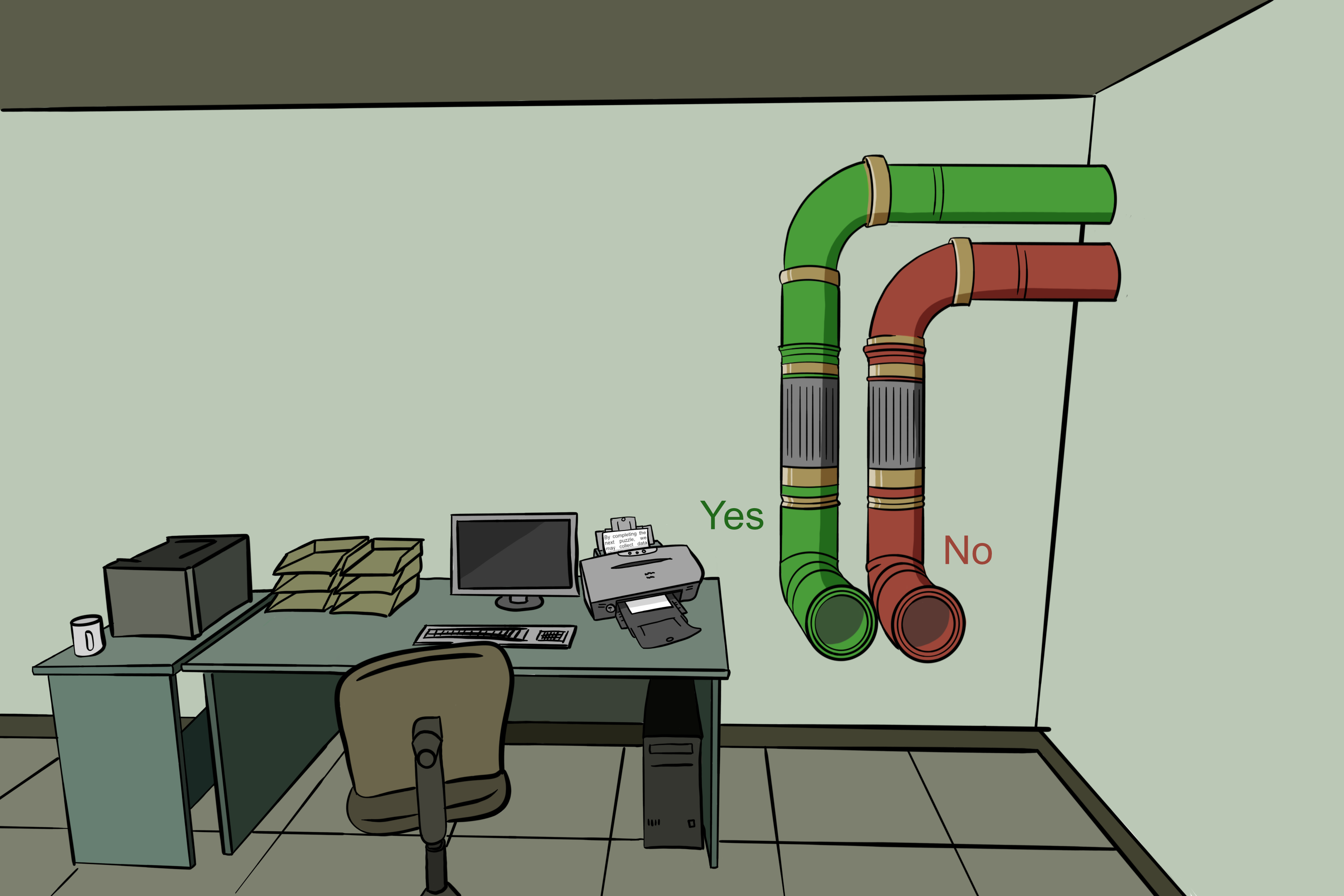}};
    \draw (-1.075, 1.72) node[inner sep=0] {\includegraphics[width=0.14\linewidth]{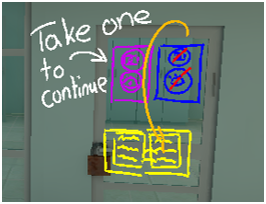}};
    \draw (-7.425, 2.3) node {\textcolor{white}{a}};
    \draw (0.2, 2.3) node {\textcolor{white}{b}};
\end{tikzpicture}
\vspace{-7mm}
  \caption{Design examples for VR data collection awareness and control interfaces developed during concept brainstorming and sketching sessions for a VR escape room game. 
  a) Players manage data collection permissions by placing accept or deny stickers in their privacy book, which is part of their game assets. The top right corner shows the original 3D sketch. b) Upon entering the room, players encounter a work desk with a privacy notice displayed on a computer screen. They can print the notice, sign it, and submit it into green or red pneumatic tubes to accept or deny data collection.
  }
  \Description{Image (a) shows a virtual hall-like setting with a door featuring a lock. Attached to the door are two papers: the first has green symbols representing camera, microphone, and eye-tracking permissions. The second shows the same symbols in red and crossed out, indicating denial. In front of the door lies a privacy book, displaying a green location symbol and a red heart rate symbol, each accompanied by a brief description. In the top right corner of this image, a lo-fi version of the scene is visible, with hand-sketched symbols and the book.
  Image (b) shows a virtual desk setup within the same environment, equipped with a computer, keyboard, and a printer printing a paper. To the right of the desk, two vertical mailing tubes are mounted into the wall, each with an open slot at desk height. The left tube is green and labeled "Yes," designed for users to insert the printed document to accept data collection. The right tube is red and labeled "No," intended for rejecting data collection.}
  \label{fig:teaser}
\end{teaserfigure}

  
\maketitle

\section{Introduction}

Sensors integrated into XR hardware bring data collection closer to the user’s physical body than previous desktop interfaces. 
As these technologies become more ubiquitous, the urgency to rethink and redesign privacy mechanisms becomes apparent~\cite{paneva2024ieeepvc}.
These challenges with privacy permissions already identified for 2D interfaces, such as smartphones~\cite{prange2024soups}, will likely become more pronounced in XR, as XR headsets have a greater variety of sensors, which partly rely on continuous data collection for functionalities such as tracking. 

In immersive VR, privacy notices and mechanisms have been directly adopted from applications designed for 2D interfaces (smartphones, PCs, tablets, etc.).
Hence, these borrowed privacy mechanisms fail to address VR environments' specific needs and challenges and do not take advantage of their unique affordances~\cite{lim2022mine}.
Existing challenges include but are not limited to (1) important information being frequently inaccessible to users due to being buried in pages of legal text~\cite{Obar2020} and (2) users typically having a limited understanding of how their privacy-related decisions impact their subsequent experience with using the application~\cite{tahaei2023stuck}.

The data collected in VR can be used not only to facilitate user interaction but also to infer sensitive information.
This can invade users' privacy, allowing third parties to obtain information users might not want to share. 
For example, telemetry and eye gaze data are indicative of users' well-being, literacy, and vision impairments~\cite{nair2023exploring}, their political views and gender~\cite{steil2019etra}, and even their identity \cite{pfeuffer2019chi,nair2023uniqueID}.
Recent studies show that users are often unaware of XR sensors' granular data collection capabilities, such as their ability to capture involuntary body signals indicative of emotions~\cite{hadan2024privacy}.

These challenges call for novel privacy concepts for VR that are (a) tailored to users and context and (b) engage users in informed decision-making regarding data privacy.
The research questions guiding our work are:
\begin{description}
    \item[RQ1] How can VR-specific privacy interfaces be designed to effectively convey data collection and provide meaningful control, considering the unique challenges and affordances of immersive environments? 
    \item[RQ2] What are the key challenges developers and designers face when creating data collection awareness and control interfaces for VR environments?
    \item[RQ3] What are some key design implications that can guide the development of effective and compliant data collection awareness and control mechanisms in VR? 
\end{description}

To address these research questions, we employed a mixed-method approach that combines hands-on design sessions with expert evaluations to systematically explore and refine privacy interfaces for VR. 
Our findings show that designers' key challenges include balancing user engagement with privacy awareness, managing the complexity of privacy information with user comprehension, and navigating trade-offs between company and user interests.
Based on insights from design sessions and expert evaluations, we derive actionable design implications, including thoughtful gamification to engage users without trivializing consent, explicit and purpose-specific consent mechanisms, and granular and modifiable consent options.

\vspace{1mm}\noindent
\textbf{Contribution Statement.} Our contributions are threefold: 
\begin{enumerate}
    \item \textbf{Low-Fidelity Design Concepts}: We present novel low-fidelity concepts for privacy interfaces in VR environments developed through participatory workshops with novice game designers and developers. 
    \item \textbf{Refined Design Concepts and Expert Evaluation}: We provide refined design concepts for informed consent in VR, evaluated in a focus group with usable privacy experts.
    \item \textbf{Design Implications}: We identify actionable design implications for developing data collection awareness and control interfaces in VR, focusing on usability, transparency, and trust. 
\end{enumerate}


\section{Related Work}

Our work draws from several strands of prior research spanning traditional privacy mechanisms designed for 2D user interfaces, data collection in XR, and approaches to incorporating privacy considerations in XR interactions.

\subsection{Data Collection Awareness}
Different approaches exist to improve user awareness of data collection.
Kelley et al.~\cite{kelley2009} initially proposed the idea of “nutrition labels” for privacy, which communicate what data is collected, how it is processed, and whom it is shared with. 
This idea has been appropriated for specific application areas by both researchers and companies, for example, as part of iOS 14. 
However, recent investigations showed that these can often be inaccurate and misleading, with discrepancies between the data disclosed in the apps' nutrition labels and actual data practices~\cite{kolling2022}. 
Privacy Badge~\cite{gisch2007} is a privacy-aware user interface for small hand-held devices designed to communicate the type of data being disclosed, when, to whom, and for what purpose. 
PriCheck~\cite{volk2022pricheck} is a browser extension providing privacy-related information about smart devices.
Privacy dashboards, such as Google's My Activity\footnote{Google MyActivity: \url{https://myactivity.google.com}}, have been introduced to help users review and manage data collected online. 
While it has been shown that privacy dashboards primarily support trust, their actual benefit in improving users' privacy remains unclear~\cite{farke2021}.
PriView~\cite{prange2021priview} is an AR application that identifies data-collecting sensors in smart home environments through visual and heat signatures, displaying the tracking space on a smartphone or a head-mounted display.

\subsection{Control Mechanisms}
Researchers investigated how users can be given control over data collection. 
The most common approach to gathering consent is notice and choice~\cite{sloan2014beyond}. 
However, notices often suffer from poor usability~\cite{rothchild2017against}.
Users generally do not know when and which data is collected. 
Furthermore, setting privacy permissions does not scale, leading users to quickly give up on managing which sensors/data each app and service should have access to~\cite{prange2024soups}. 

Researchers proposed visually appealing privacy notices~\cite{kitkowska2020} and
notices considering timing, functionality, and modality~\cite{feng2021}. 
Morrison et al.~\cite{morrison2014improving} showed that users confronted with visual representations of collected data by smartphone apps became more concerned and often stopped app usage sooner.
Studies on user behavior when granting permissions explored visual cues to support decision-making~\cite{rajivan2016influence}, and statistical insight to improve understanding of the implications of consenting to permissions~\cite{kraus2014}.
King et al.~\cite{king2011} showed that designing privacy choices to be personal and concrete supports users' privacy decisions. 
Liu et al.~\cite{liu2016} developed a privacy assistant that learned user preferences in privacy settings, with almost 80\% of the recommendations adopted with only small post hoc changes. 
Momen et al.~\cite{momen2020} introduced partial consent, adding an indecisive state that allows users to evaluate data services before providing full consent. 
Similarly, an approach beyond binary permission options was proposed as a fine-grained privacy permission slider for AR applications~\cite{abraham2024}.

\subsection{Data Collection in XR}

XR headsets utilize an array of sensors (including many cameras) and bring them closer to the human body, enabling sensitive data to be captured, processed, and shared with third parties.
State-of-the-art headsets already provide access to behavioral data (e.g., hand and body motion, eye gaze), physiological data (e.g., electroencephalography, heart rate), contextual data (e.g., size of tracking space, bystanders), and device specifications~\cite{nair2023exploring, garrido2024sok}. 
Such data allow one to infer information about user demographics (e.g., age, gender, handedness), health and well-being (e.g., reaction times, fitness level), impairments (e.g., visual or motor impairment), and emotional state~\cite{steil2019etra, nair2023exploring, Tabbaa2022}.
Moreover, XR users often store online account credentials and financial details, including bank account and credit card numbers, directly on their devices~\cite{Zhang2018, Zhu2020}.
While this data benefits XR users (e.g., novel functionalities) and stakeholders (e.g., target ads), it also has severe privacy implications for users~\cite{guzman2019}.
A comparative analysis of the network traffic from VR applications by Trimananda et al.~\cite{trimananda2022ovrseen} found that approximately $70\%$ of the data transmissions were not adequately disclosed in the privacy policies.
Adams et al.~\cite{adams2018ethics} proposed a code of ethics advocating for secure protocols, transparency in data collection practices, and obtaining user permission each time data is collected. 
However, practical guidance on implementing these principles in practice is still lacking.


\subsection{Designing Usable and Privacy-Preserving XR Interactions}

The design of XR interactions often prioritizes usability and feasibility at the expense of security and privacy. Recent work addresses this gap by embedding privacy considerations into development processes, aligning the expertise of XR developers and privacy specialists. For example, Rajaram et al.~\cite{Rajaram2023} employed scenario-based threat modeling in multi-user AR to balance competing goals of usability, feasibility, and security/privacy. Ruth et al.~\cite{Ruth2019} proposed design goals that accommodate security needs while maintaining functionality, exemplified by a content-sharing prototype for multi-user AR. Tools like \textit{SecSpace}~\cite{Reilly2014} and \textit{Reframe}~\cite{rajaram2023reframe} further help developers identify security and privacy risks early in AR/MR design. However, these frameworks primarily focus on collaborative AR/MR environments, leaving VR-specific threats and development patterns less studied.

\subsection{Summary}

Privacy challenges in XR are distinct due to the variety and proximity of sensors to the human body, capturing sensitive data beyond what is common in desktops, smartphones, and smart homes.
While previous studies on 2D interfaces propose approaches for data collection awareness and control, such as privacy labels or assistants, these often fail to provide users with a sufficient understanding of when and why data is collected or the implications of their choices. 
XR devices, in particular VR headsets, present additional complexities as they collect not only environmental data but also physiological and behavioral data. This includes eye-tracking, hand movements, and body posture, revealing sensitive information such as emotions, health conditions, or even identity~\cite{nair2023exploring, nair2023uniqueID, pfeuffer2019chi}. These challenges, acknowledged by researchers and industry~\cite{adams2018ethics, abraham2022implications, hadan2024privacy, ttcLabsXR}, highlight the need for VR-specific privacy interfaces to effectively inform and enable meaningful control over VR data (\textbf{RQ1}).

Tools like~\textit{SecSpace}~\cite{Reilly2014} and \textit{Reframe}~\cite{rajaram2023reframe} facilitate integrating privacy considerations in multi-user MR and AR interactions. Yet, there is limited systematic research on the practical challenges practitioners face in designing privacy interfaces for immersive virtual environments. 
Understanding these challenges is critical to bridging the gap between user needs and constraints in hands-on implementation, motivating our focus on the obstacles designers and developers face during the development process for VR (\textbf{RQ2}).
Prior AR research indicates that incorporating privacy considerations often requires balancing competing design goals, such as usability, feasibility, and privacy~\cite{Rajaram2023}. 
However, comprehensive guidelines for designing effective and compliant VR privacy interfaces are still lacking.
This gap highlights the need for actionable design implications to guide developers and designers in creating robust privacy mechanisms in VR, balancing usability, transparency, and compliance (\textbf{RQ3}).

\begin{figure*}[t]
    \centering
   \includegraphics[width=\linewidth]{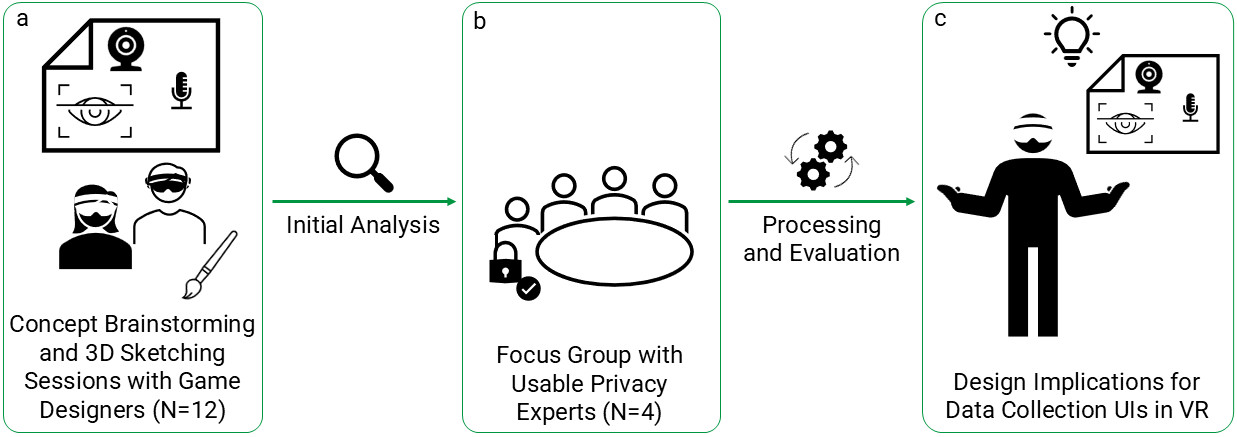}
\vspace{-6mm}
  \caption{Overview of the design process for VR data collection awareness and control interfaces. a) Novice game designers and developers ideate and sketch innovative solutions tailored to 3D virtual environments during concept brainstorming and sketching sessions. b) The generated design concepts are analyzed and discussed in a focus group with usable privacy experts to assess effectiveness and feasibility. c) Insights from the design and evaluation stages inform a set of design implications. 
  }
  \Description{The figure illustrates a three-part overview of the design process.
Bubble (a) features icons of three people wearing Head-Mounted Displays (HMDs) with a paintbrush symbol, indicating their role in developing creative solutions. Above them is a page displaying symbols for camera, microphone, and eye-tracking, highlighting the focus on data collection elements during the design phase. An arrow with a magnifying glass icon points towards the next stage.
Bubble (b) shows three people seated at a table, accompanied by a lock icon with a tick, symbolizing their role in reviewing and assessing the privacy aspects of the designs. An arrow with a processing icon above it leads to the final stage.
Bubble (c) shows a person interacting with an XR system, with a privacy notice and a lightbulb icon above them, representing the application of insights gained from the previous stages to create effective and user-friendly privacy controls.}
  \label{fig:methodology}
\end{figure*}

\section{Research Approach}

We employ a mixed-method approach consisting of concept brainstorming and sketching sessions with novice game designers and developers (\autoref{fig:methodology}a) and a focus group with usable privacy experts (\autoref{fig:methodology}b) to generate and refine design implications for data collection awareness and control interfaces in VR (\autoref{fig:methodology}c).

This approach draws from similar methods in prior research, such as the user-centered design process employed by Fassl et al.~\cite{fassl2021exploring} for developing authentication ceremonies for instant messaging, and the involvement of novice AR designers and security and privacy experts in the design and evaluation of an AR storyboarding tool with threat modeling by Rajaram et al.~\cite{rajaram2023reframe}.
Given VR's predominant use in entertainment, particularly gaming (approximately 70\% of VR consumers\footnote{\url{https://www.demandsage.com/virtual-reality-statistics/}}), where in-game behavior can reveal sensitive information about players such as their interests, emotions, and skills ~\cite{kroeger2023surveilling}, we chose to use the open source \emph{MetaData} escape room game by Nair et al.~\cite{nair2023exploring} as a sandbox environment for our study.
As an example of an adversarial VR game constructed to harvest user data, it is particularly suitable for exploring data collection awareness and control interventions due to its high ecological validity and various examples of collected data and inferred user information.

In our study, novice game designers and developers developed intuitive and engaging data collection awareness and control VR interfaces, individually or in groups (depending on availability), during think-aloud brainstorming and sketching sessions using a virtual 3D sketching tool.
The 3D sketching tool allowed participants to prototype directly within the VR scene, providing a better understanding of placement, proportion, and effective use of the virtual 3D space. 
Studies have shown that 3D sketching can be more stimulating and engaging, especially in the early design process~\cite{houzangbe2022}, as it promotes physical and perceptual action and enhances flexible cognitive thinking more than traditional 2D methods~\cite{lee2019design}.
These live sessions allowed us to gain immediate insight into the topics and challenges that naturally emerged during the design process. 
Following the brainstorming and sketching sessions, selected design concepts were refined and re-sketched for further evaluation in a focus group with usable privacy experts.
We made audio recordings during the design sessions and the focus group to capture participants’ discussions, observations, and insights. 
We transcribed and analyzed the recordings using thematic analysis.
Lastly, using the combined insights from the hands-on design sessions and the expert feedback, we discuss different strategies for effectively integrating privacy-related information into immersive VR applications and synthesize actionable design implications.

\section{Concept Brainstorming and Sketching Sessions}
This section describes the process of generating privacy interface concepts for VR environments, detailing the apparatus, procedure, participants, and data analysis.

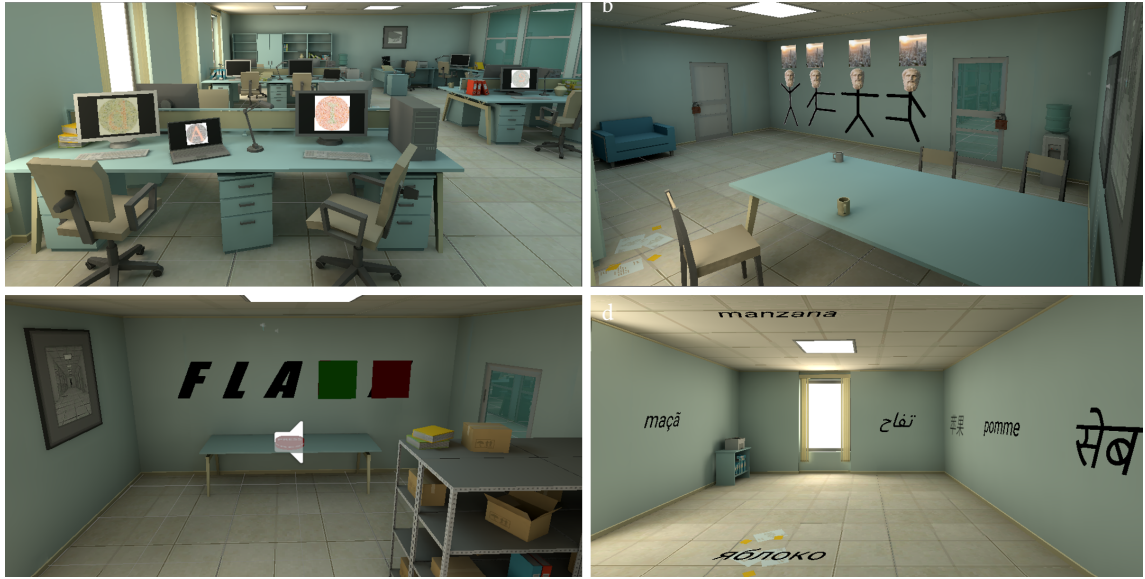
\begin{figure}[t]
    \centering
    \begin{tikzpicture}
    \draw (0, 0) node[inner sep=0] {    \includegraphics[width=\linewidth]{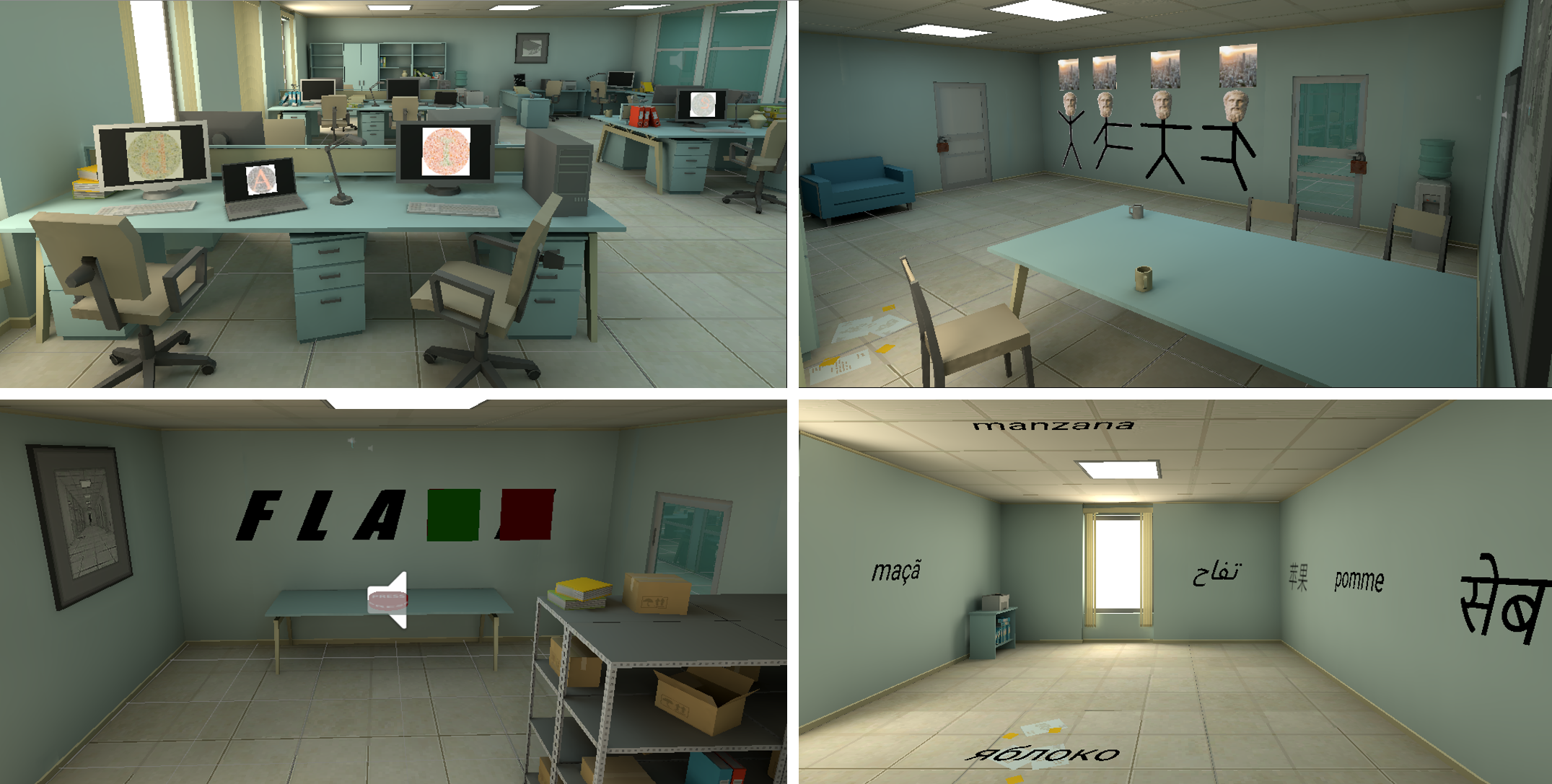}};
    \draw (-7.7, 3.8) node {\textcolor{white}{a}};
    \draw (0.45, 3.8) node {\textcolor{white}{b}};
    \draw (-7.7, -0.3) node {\textcolor{white}{c}};
    \draw (0.45, -0.3) node {\textcolor{white}{d}};
\end{tikzpicture}
    \caption{Puzzle rooms from~\cite{nair2023exploring}. a) The monitors display different letters on Ishihara color test plates. Depending on the word the user can read, color blindness can be inferred. b) The user has to mimic the poses displayed on the wall to reveal the password, enabling measurement of their wingspan. c) When the panels on the wall turn from red to green, the user has to press the button on the table to uncover the letter of the password, revealing their reaction time. d) The password is written in different foreign languages around the room; the direction of the user's gaze reveals which language(s) they recognize.
    }
    \label{fig:PuzzleRooms}
    \Description{The figure presents four images of a virtual office environment.
Image (a) displays an open office setting with multiple monitors showing letters on various color plates, resembling an eye test. 
Image (b) shows a virtual meeting room with a table, chairs, a sofa, and a water cooler. On the wall, there are illustrations of four stick figures demonstrating different poses: both arms up, left arm and leg up, arms stretched, and right arm and leg up.
Image (c) depicts an office storage room with shelves of boxes along the right wall. In the center, there is a table with a prominent red button. Above the button are three letters and green and red panels.
Image (d) shows an empty office room with words in various languages written on the walls, floor, and ceiling.}
\end{figure}

\subsection{Apparatus}

The sketching sessions were conducted using an HTC Vive Pro HMD. 
We used the open-source virtual environment by Nair et al.~\cite{nair2023exploring}.
To keep the duration of the full design sessions within a reasonable time frame and avoid participant fatigue, we presented four rooms to the participants, each measuring different attributes: color blindness, wingspan, reaction time, and foreign language knowledge (see Figure~\ref{fig:PuzzleRooms}).
From these, participants selected two for their design development.   
We built a 3D sketching application on top of the Unity game engine to allow users to prototype directly within the virtual scene.
The sketching application allowed users to draw freehand lines in 3D, 
resize and rotate the sketch, and perform undo actions.  
Users performed the 3D sketching using hand-held HTC Vive controllers.

\subsection{Procedure}
We began by explaining the study’s purpose and procedures, then obtained signed consent for voluntary participation and audio recording. Participants completed a demographics questionnaire and the IUIPC questionnaire~\cite{malhotra2004internet}. Next, we familiarized them with the VR escape rooms and the 3D sketching tool, allowing time for practice. Once comfortable, they moved on to the main task: creating low-fidelity prototypes of data collection awareness and control interfaces within different escape rooms. They first brainstormed ideas on paper, then selected promising concepts for implementation in VR, using think-aloud protocols to articulate their design decisions as they sketched. Afterward, participants filled out the IUIPC questionnaire again and participated in a semi-structured interview. The interview focused on the specifics of their designs, the challenges encountered, and their overall reflections on privacy considerations in game development. The interview protocol is provided in \autoref{semi-structure_interview}.


\subsection{Participants}
We employed convenience sampling followed by snowball sampling to recruit 12 participants (2 females, 9 males, 1 non-binary), aged between 21 and 26 years (mean age 23.92, SD 1.56). 
Participants were drawn from a cohort of students enrolled in Computer Games and Media Studies programs, all with prior experience in game design. 
Eligibility criteria included having completed at least one course related to game design, having worked on a game design-related project, or having professional experience in the game design industry. 
Ethical approval was obtained from the Ethical Committee of the University of Bayreuth.
All participants were reimbursed for their participation.
The study lasted approximately 90 minutes.

\subsection{Data Analysis}
The collected data included audio recordings of the think-aloud prototyping sessions, 3D sketches of conceptual prototypes, and audio recordings of the concluding semi-structured interviews. 
We collected and analyzed 395 minutes of audio recordings in total, with an average of 56.43 minutes (SD 15.59) per design session.
We transcribed the audio recordings using MAXQDA transcription software and reviewed the sketches to better understand their structure and design intentions. 
Two researchers independently open-coded the transcript from one randomly selected prototyping session, accounting for approximately 15\% of the data, and met to discuss the initial codes and form a common codebook.
Three researchers then used the common codebook to code the remaining prototyping sessions.
The three researchers compared and discussed codes, resolving disagreements in a final meeting to finalize the codebook.
The codes of the final codebook were then compared and grouped to identify common themes. 
The same process was applied to the transcripts of the semi-structured interviews.

\section{Focus Group}

We conducted a focus group with usable privacy experts to evaluate some of the data collection design concepts generated during the prototyping and sketching sessions. 
We also explored challenges and opportunities in developing usable privacy mechanisms for immersive VR.

\subsection{Participants}
The focus group consisted of four participants (3 males, 1 female), aged 24--31 (mean age 28.25, SD 2.68), recruited from the authors' academic research network.
All focus group experts pursue or have completed a PhD in usable security and privacy. 
One of the experts (E3) also has a 3 year experience in XR research and development.

\subsection{Procedure}

The focus group was structured into two parts. 
The first part involved a hands-on SWOT analysis of the designs.
In the second part, the experts used the insights from the design analysis to discuss potential challenges and trade-offs and explore design implications for privacy-aware data collection interfaces in VR. 

The focus group began with a brief introduction to the study's context, followed by a description of the concepts and presentation of the sketches. 
The experts first analyzed each design concept individually using the SWOT matrix, followed by a group discussion. 
Drawing on the insights from these discussions, the experts were asked to draft design recommendations for usable privacy mechanisms in VR on paper, which were then collectively discussed. 
In the next phase, we wrote challenges identified during the design sessions with the novice game developers and designers on separate post-its and thematically organized them on a whiteboard.
We then invited the experts to comment, reorganize, and add any aspects that might be missing in the overview, providing further recommendations specific to the challenges encountered by novice game developers and designers. 
The session concluded with a wrap-up and final comments. 
The focus group session lasted 90 minutes. 

\subsection{Data Analysis}

We recorded approximately 90 minutes of audio, transcribed using MAXQDA transcription software. 
Two researchers independently conducted open coding in MAXQDA, then met to discuss the initial codes and collaboratively developed a common codebook. 
The codes were analyzed and grouped into overarching themes. 
The individual SWOT matrices were compiled into a joint table to identify patterns and points of agreement/disagreement among the experts.

\section{Results}

\begin{figure*}[t]
    \centering
    \includegraphics[width=\linewidth]{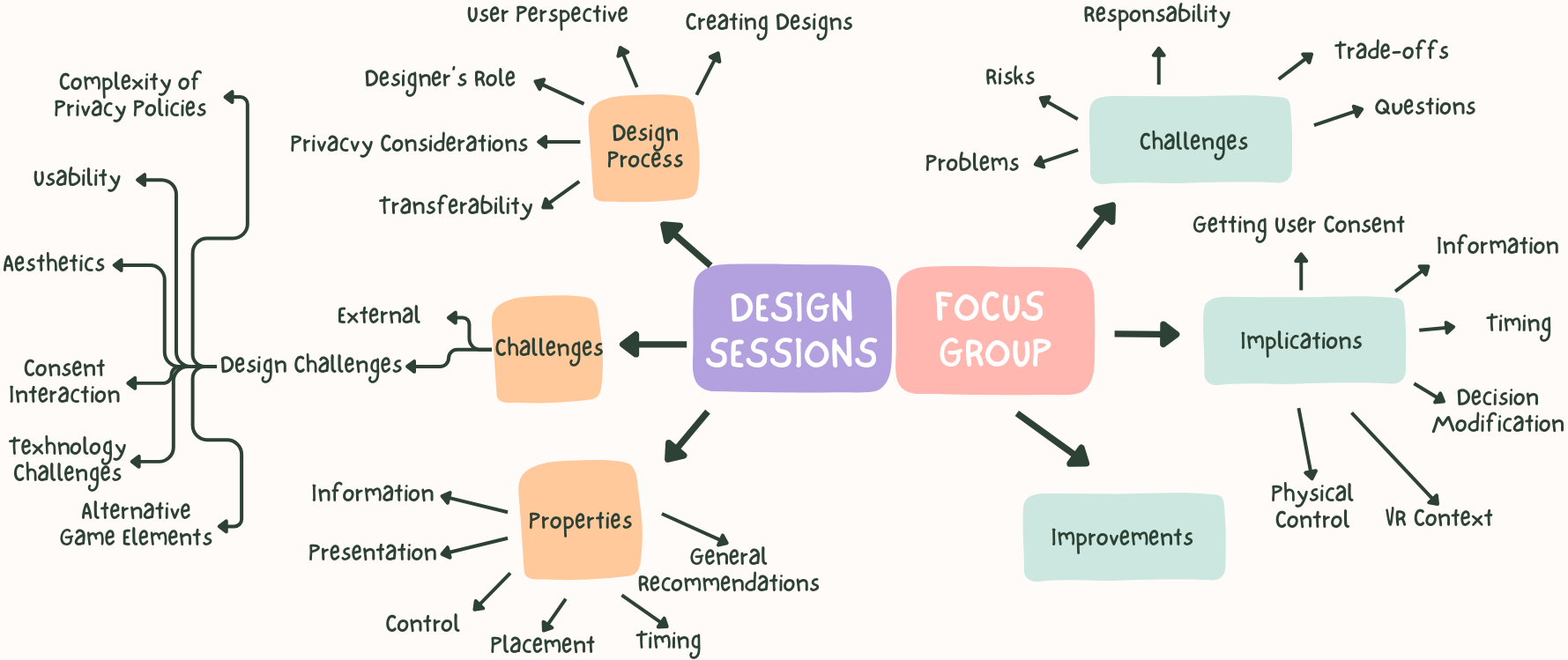}
    \caption{Overview of themes emerging from the thematic analysis of the design sessions and the focus group. }
  \Description{The figure is a mind map illustrating key themes and topics related to "Design Sessions" and "Focus Group". The map is divided into several interconnected sections with nodes and sub-nodes organized around the central themes. The three main themes from the design sessions are Design Process, Challenges, and Properties. The Design Process theme encompasses aspects such as the user perspective, the designer's role, privacy considerations, the transferability of designs, and the process of creating designs. The Challenges theme is divided into external challenges and specific design challenges, including the complexity of privacy policies, usability, aesthetics, consent interaction, technology challenges, and alternative game elements. The Properties theme encompasses information, presentation, control, placement, timing, and general recommendations. The three main themes of the focus group are: Challenges, Recommendations, and Improvements. The Focus Group Challenges theme includes responsibilities, risks, problems, trade-offs, and questions. The Recommendations theme includes getting consent, information, timing, decision modification, XR context, and physical control.
  }
  \label{fig:themes}
\end{figure*}

We first outline the design concepts from the concept brainstorming and prototyping sessions, followed by evaluations with privacy experts. 
Through thematic analysis, we identify key design challenges and implications for VR data collection awareness and control interfaces.
The following sections detail our findings, with an overview of all the themes provided in~\autoref{fig:themes}.

\subsection{Results Design Session}

We provide an overview of the generated designs, an insight into the design process, and the challenges that the novice game developers and designers faced.

\subsubsection{Designs}
We briefly describe the different design concepts developed during the design sessions. 
A summary of all designs can be found in \autoref{table:designs_all}.
Several approaches focused on informed consent mechanisms, such as displaying privacy notices on computer screens for players to sign and submit through pneumatic tubes to accept or deny data collection (Design D2), attaching privacy notices to the player's avatar to accept or disposing of them in virtual trash cans to deny (Design D4), and placing stickers in a privacy book to manage data collection permissions (Design D17).
Other designs explored alternative game mechanics to minimize data collection, such as manipulating a mannequin (Designs D3 and D6) or drawing on a whiteboard (Design D13) instead of using body tracking, which would otherwise capture telemetry data, such as height and wingspan.
A general alternative for all data collection tasks included hiding password letters within the virtual environment, requiring players to search for them instead, thereby eliminating direct data capture from specific interactions (Designs D11 and D14).
Awareness-raising designs included using visible cameras with indicator lights to signal active user tracking (Designs D9 and D15) and interactive companions, such as a robot or other game character, that inform players about data collection (Design D16).

\begin{table}[]
\centering
\caption{
Summary of design concepts developed during the brainstorming and sketching sessions, exploring different approaches to data collection awareness and control in a VR escape room game.
These include informed consent mechanisms, alternative game mechanics to minimize data collection, and awareness-raising strategies. The table includes participant ID, attribute, type, design number, and concept description.
Designs marked with \textbf{*} were selected for further analysis with privacy experts in the focus group. Abbreviations: RT = Reaction Time, W = Wingspan, L = Language.
}
\footnotesize

 \begin{tabularx}{\textwidth}{|c c l l X|} 
 \hline
 \textbf{ID} & \textbf{Attribute} & \textbf{Type} & \textbf{No.} & \textbf{Description}  \\ 
 \hline\hline
P1 & RT & Awareness & D1 & \faBug~The player plays a minigame where they swat virtual bugs to raise awareness about collecting reaction time data. \\ 
P2-4 & All & Informed Consent  & D2* & \faPaperPlane~A privacy notice is displayed on a computer screen, which the player can sign by typing their name on the keyboard and then print. To accept or decline, players place the printed document into one of two labeled pneumatic tubes on the wall: "Yes" to accept and "No" to deny data collection. A "woosh" sound confirms the document has been sent.\\ 
& W & Alternative &  D3 &  \faChild~A mannequin is placed in the room, which the player can manipulate to display the poses needed to reveal the password. \\
P5-6 & All & Informed Consent & D4 & \faTrash*[regular]~There is a piece of paper in each room, briefly informing players about data collection. Players can provide consent by attaching the paper to their avatar, similar to collecting items in games, or deny it by simply throwing it into a trash can.\\ 
P7 &  W & Informed Consent & D5 &\faStreetView~The player stands in a body scanner marked by a circle on the floor, with a projector-like device above. A privacy notice appears on a display. To accept data collection, the player must mimic a pose projected by the device; otherwise, they are teleported to the next room.
\\ 
& W & Alternative & D6 &  \faChild~The room contains four mannequins that the player can adjust to match the required poses.\\
& W & Alternative &  D7 & \faDotCircle[regular]~A stack of interactive balls is placed on a table to serve as markers. The players must position these markers in mid-air to replicate the poses on the wall, revealing the password.  \\ 
 & RT & Alternative & D8 &  \faPuzzlePiece~The player must solve a logic pattern puzzle instead.
 \\ 
P8 &  All & Awareness & D9 &  \faVideo~There are multiple cameras mounted in each room, and the red indicator light will be turned on to signify ongoing data collection to the player.  \\ 
& L &  Alternative & D10 & \faGlobeAfrica~The password is written on an interactive globe in the language predominantly spoken in each respective country. \\ 
& All & Alternative & D11 &  \faObjectGroup~The letters of the password are hidden in the room; the player must search for them to find the password.\\ 
P9 & All & Informed Consent & D12* &  \faChalkboard~The privacy notice is written on a whiteboard. The player can exercise control by using one of the two erasers labeled "accept" or "deny".\\ 
& W & Alternative & D13 &   \faPaintBrush~The player must draw the different poses on the whiteboard. \\
& All & Alternative & D14 &   \faObjectGroup~The letters of the password are blended within other objects, requiring the player to search and identify them. \\
P10-12 & All & Awareness & D15 & \faVideo~A camera with a blinking red light indicates data collection.\\ 
&  All & Awareness & D16 &  \faRobot~A privacy companion, e.g., a robot, accompanies the player and provides them with information about data collection.\\ 
& All & Informed Consent & D17* & \faBookOpen~On the door of each room, there are "accept" or "deny" stickers for each data collection type. The player exercises control by placing these stickers in their data collection book.\\ 
 \hline
 \end{tabularx}
\label{table:designs_all}
\end{table}

\subsubsection{Design Process}

We identified five categories that group the codes in this theme.

\emph{Creating Designs} describes the designers’ feedback regarding the design activity and their experience when creating the designs in a 3D environment. They mentioned that "it was very fun, (...) like a cool way of puzzle solving" (P2) while being "different than drawing in 2D" (P1), which took time to adjust to (P1, P4).

\emph{User Perspective} and \emph{Designer’s Role} reflect on how considering the user perspective informed the designs and how designers and developers perceived their roles and responsibilities regarding privacy.
P6 mentioned that "if a game would sell [the data] to companies for like advertisement (...) then it is [a] concern, but to improve the game I think it is all right", and P1 added that it is "a compromise you make with yourself" when sharing data to use an application.
Regarding their role as developers, participants mentioned that they "have to be aware [of the data processing] while designing the games" (P2), and the trust users place in developers, who in their role have higher awareness of these issues (P1).

In \emph{Privacy Considerations}, we grouped the privacy aspects designers saw as important for their design. 
Multiple participants mentioned that privacy considerations are not something they usually think about (P1, P2, P7, P9). P1 and P7 also noted that the importance of these considerations varies with the type of game they develop and the company's size and interests.
The final category, \emph{Transferability}, contains the designers’ thoughts about how well their designs adapt to other contexts.
Most designs were described as requiring adaption, for example, to reflect very specific scenarios (P2, P5, P7) or a different number of privacy choices that must be covered (P7). P5 also thought about "mak[ing the design] more basic" to fit multiple applications more easily.

\subsubsection{Challenges}
 We identified two main themes related to the challenges designers faced: \emph{External Challenges} and \emph{Design Challenges}.
\emph{External Challenges} include the influence of company structures and legal requirements. P3 raised the question of which perspective is taken during the development: "Are we the company or are we pro people's data?", hinting at a possible conflict between user and company interests. While participants were aware that privacy mechanisms had to be legally correct (P3, P5), they also noted a lack of confidence in determining a legally appropriate privacy-preserving design (P3).

\emph{Design Challenges} contains different aspects to be considered during the design, such as \emph{Complexity of Privacy Policies, Usability, Aesthetics, Consent Interaction, Technology Challenges,} and designing\emph{Alternative Game Elements}.

\begin{description}
    \item[Complexity of Privacy Policies] Several factors contribute to the complexity of privacy policies, especially in VR, where a lot of privacy decisions have to be considered. Participants P4, P5, and P10 highlighted the importance of avoiding lengthy, repetitive interactions while still covering the different areas of data collection within the environment and the different rooms: "If only one type of data is collected, it would be easy to have two doors. But with more types of data it is difficult" (P10).
    
    Reading a lot of text also falls into these repetitive tasks. While text is mentioned as an easy and established tool to document relevant information (P5), especially long blocks of text can lead users who "just want to play the game [to] just accept because it is easier" (P5), which in turn does not result in informed consent. When a lot of information about the data collection is shared with users, the designers also mentioned the risk of "frightening the user" (P11) by bringing the privacy aspects to their attention in the first place.
    
    Another aspect adding complexity is related to the information inferrable from the collected data. While this inferred information is often less obvious than the basic sensor data, it can reveal sensitive information (e.g., gender). Informed consent should include awareness of the consequences of data collection and also cover the inferred data (P9, P10). Depending on the combination of different sensor data or the recorded user behavior, the designers saw getting consent for all possible inferred data as challenging.
    
    \item[Usability] Participants noted that the action to accept or decline data collection must be intuitive and accessible to users with disabilities (P5), for example, by not relying on hearing abilities or having the user say something to interact with the privacy mechanisms as the only option. P8 also highlighted the need to communicate the current tracking state to the user so that they are aware of it at all times.
    
    \item[Aesthetics] Participants highlighted that a fitting aesthetic design of the privacy notice could help maintain immersion in the VR environment. This can also be supported by using immersive game elements as an interaction method for the privacy decision (P2, P7, P9, P10) and by choosing a format that is appealing to the user (P10, P11) to prevent disrupting the game experience.
    
    \item[Consent Interaction] Acquiring the user’s explicit consent is critical. P5 highlighted the need for a fallback when users neither accept nor reject permissions, such as by ignoring the prompt. P8 proposed blocking other game elements until a decision is made, thus enforcing a response. In VR contexts, P5 underscored the importance of adapting consent interactions to each specific setting, with P6 arguing that general solutions are often insufficient and may require custom designs—albeit at a higher development cost. Finally, P5 questioned whether reducing text for a more immersive experience might reduce clarity, risking unintended user choices.
    
    \item[Technology Challenges] Participants highlighted several technology-related challenges in VR environments.
    Due to the sensors used in XR devices, constant data collection is a relevant aspect brought up by P10. 
    Multiple participants (P2, P5, P6, P9-11) stressed the importance of obtaining consent before the tracking begins.
    P5 and P6 discussed the importance of persistent privacy choices across multiple sessions to improve usability and prevent repetitive interactions. Another challenge arises when users want to revoke consent after initially agreeing to data collection. P5 mentioned a potential conflict if the data had already been shared with third parties or combined with other data points to infer further information about the user, making reversal difficult.
    
    \item[Alternative Game Elements] Several designers (P2, P5, P8-12) brought up the importance of alternative game elements to allow users to continue using the application even if they do not (fully) consent to sharing their data. P8 also mentioned that the alternative game elements should be easy to understand and ideally as playful as the original interaction.
\end{description}

In the sessions, designers also highlighted current limitations of privacy mechanisms for 2D environments, including deceptive designs (P3), walls of text (P3, P8, P9), lack of transparency (P2) and user-friendliness (P2, P8). 

After completing the brainstorming and sketching sessions, we observed an average increase in the IUIPC scores of the novice game developers and designers across all three dimensions.
The average control score increased from 5.89 (SD 0.74) to 6.28 (SD 0.77), awareness from 6.24 (SD 0.56) to 6.36 (SD 0.85), and collection from 4.81 (SD 1.05) to 5.04 (SD 1.51). 
An overview is provided in~\autoref{table:IUIPC}.

\begin{table}[t]
\centering
\caption{Overview of scores for the IUIPC dimensions of Control, Awareness, and Collection before and after the brainstorming and sketching sessions.}
\begin{tabular}{|l|llll|}
 \hline
\multirow{2}*{\textbf{IUIPC dimension}} &  \multicolumn{2}{c}{\textbf{Pre-study}} & \multicolumn{2}{c|}{\textbf{Post-study}} \\
& Mean (SD) & Range & Mean (SD) & Range  \\
  \hline  \hline
Control & 5.89 (0.74) & (4 - 6.67) & 6.28 (0.77)  &  (4 - 7)  \\
Awareness  & 6.24 (0.56) & (4.67 - 6.67) & 6.36 (0.85) & (3.67 - 7) \\
Collection   & 4.81 (1.05) & (2 - 6.75) & 5.04 (1.51)  &  (2 - 7)  \\
  \hline
\end{tabular}
\label{table:IUIPC}
\end{table}

\subsubsection{Properties of the Privacy Mechanisms}

This theme contains the properties of the designed privacy mechanisms in VR identified by the participants, alongside general design recommendations. 
The categories identified within this theme are \emph{Information, Presentation, Control, Placement, Timing} and \emph{General Recommendations}. 

\begin{description}
    \item[Information] addresses the content of the notice and the details users require to give informed consent. Participants stressed that notices should specify which data types are collected (e.g., height, wingspan) and what can be inferred from them (e.g., gender), along with explaining why these data are needed. They also noted the importance of conveying the potential consequences of a user’s privacy choices, including how gameplay might change. To manage the amount of information displayed, some suggested providing a link to a more detailed policy. Additionally, participants recommended offering an overview of collected data after consent, enabling users to review what has been captured.
    
    \item[Presentation] Participants discussed different ways to present privacy notices, including the use of text (P5, P8) and symbols (P10). They also emphasized the need to keep the amount of presented information manageable for the user (P10), adapt it to previous choices, and keep the interaction serious but fun to make it engaging while still fulfilling the purpose of the notice as a privacy control mechanism.

    \item[Control] Participants emphasized the importance of obtaining explicit user consent: "There needs to be the step of confirmation" (P11). Consent should be given through an \emph{opt-in} mechanism rather than default collection. The control options should include reviewing collected data (P9), deleting it, and revocation.

    \item[Placement] The designers explored different placement options for privacy notices, including integrating the notice with the VR environment and the existing game elements (P9). The notice should be reachable for the user (P4) to make it easy for them to interact with it.
    Most designs placed the notice in front or at the start of the room (P5, P10, P11), so either located where the user would enter (P11) or get teleported to, ensuring they face it directly and get prompted to interact (P9). Another idea was to position another part of the notice at the end of the room, informing users about the data collected during that section of the game: "At the end of the game (...) you can look at [your inventory] and see the data" (P10).
    
    \item[Timing] Beyond placement, designers suggested that certain actions could trigger the start of the interaction or present symbols to inform the user about which actions cause their data to be collected and when for example, using "notification pop-ups that you got some new [data] entry" (P11).
    
    \item[General Recommendations]  Participants emphasized the importance of continuously informing users about ongoing data collection (P8). They recommended offering detailed explanations on demand to avoid overwhelming users with lengthy text, while still providing access to in-depth information when needed.
\end{description}

\subsection{Results Focus Group}

This section details the refinement of the design concepts evaluated in the focus group. We discuss additional challenges associated with privacy-aware design and present design recommendations provided by the experts based on their evaluations of the three selected design concepts. 
 
\begin{figure*}[t]
    \centering
    \begin{tikzpicture}
    \draw (0, 0) node[inner sep=0] {
    \includegraphics[width=0.5\linewidth]{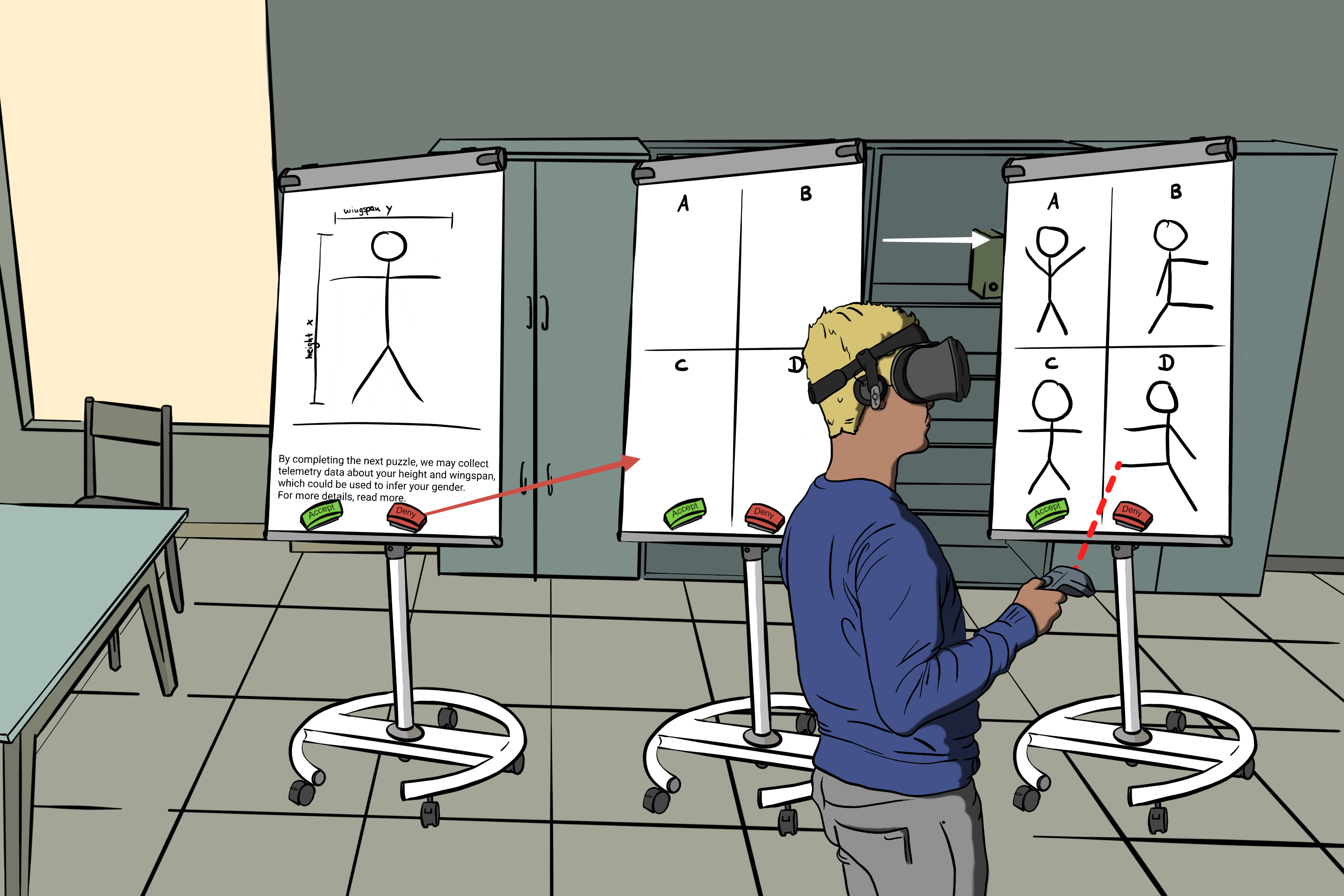}
    \includegraphics[width=0.5\linewidth]{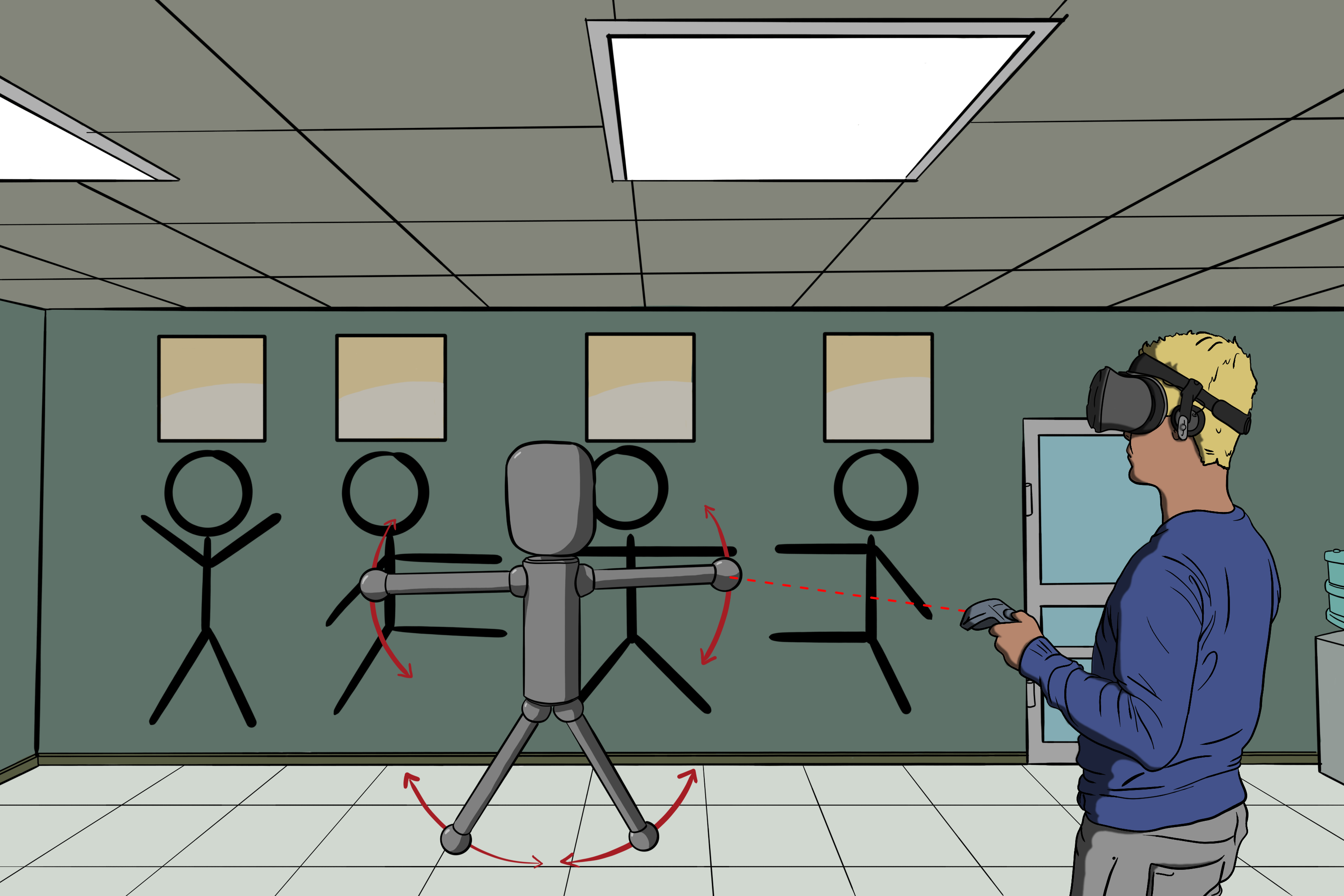}};
    \draw (-7.8, 2.4) node {\textcolor{black}{a}};
    \draw (0.2, 2.4) node {\textcolor{black}{b}};
\end{tikzpicture}
  \caption{Privacy control interfaces in the telemetry escape room, where users have to perform a set of four poses to uncover a password on the wall, thereby revealing their height and wingspan. (a) The user interacts with a whiteboard displaying data collection information, using green and red erasers to accept or deny permissions. Denying triggers an alternative task where players draw poses instead of performing them physically. (b) The user can manipulate a virtual 3D mannequin to perform poses, providing a less data-intensive approach to completing the task and uncovering the room’s password.
  }
  \Description{(a) The image depicts a person wearing a VR headset and holding a controller, engaging with a series of flip charts in a room resembling an office or classroom with tiled floors and cabinets. The individual is focused on the charts, interacting with them using the VR controller. The left flip chart shows a stick figure with measurements for height and wingspan, accompanied by text indicating that completing the next puzzle may collect telemetry data that could infer the user’s gender, with options to “Accept” or “Deny” displayed below. The center chart shows four empty quadrants labeled A, B, C, and D, while the right chart contains the same quadrants filled with stick figure poses, featuring "Accept" and "Deny" buttons. The overall setup highlights an interactive experience where the user’s actions within the VR environment might involve decisions about data collection, underscoring the importance of clear virtual feedback and accessible options for interaction. (b) The image shows a person wearing a VR headset and holding a controller, interacting with a digital figure in an office with tiled floors and ceiling panels. In the center of the room stands a gray mannequin in a T-pose, with red arrows around its arms, legs, and body indicating movement or adjustments, likely controlled by the VR user. On the wall behind the figure, there are several stick figure drawings in various poses, serving as reference or target positions for the VR activity. The person, focused on the VR experience, appears to be manipulating the robotic figure to match these poses using their controller.}
  \label{fig:wingspan_room}
\end{figure*}

\subsubsection{Design Reiteration}
\label{focusgroup_design_reiteration}
After the brainstorming and sketching sessions, we reviewed all design concepts in~\autoref{table:designs_all} of type \textit{Informed Consent}, i.e., the ones that propose both an awareness and a control mechanism. We excluded design D5, because its control mechanism already required collecting user (telemetry) data. D2 and D4 were similar -- representing consent through an embodied user interaction with a virtual piece of paper. We proceeded with D2 for further refinement.
Finally, D2, D12, and D17 were re-sketched with added detail to ensure they accurately represented the proposed data collection awareness and control mechanisms. 
We provide a brief description of the refined concepts:
\begin{description}
    \item[Design 1] Accept and deny stickers are displayed at the door of each room for each type of data being collected (\autoref{fig:teaser}a). 
    The player exercises control by placing these stickers in their privacy book, which is part of their assets. 
    This book allows players to manage and adjust their data permissions by adding or removing stickers, effectively accepting or denying data collection as needed.
    \item[Design 2] The player is teleported into the room, facing a work desk. 
    On the desk, a computer screen displays a privacy notice outlining the data that will be collected. 
    They can sign it using the keyboard before using the printer next to the monitor to print the notice.
    Next to the desk, two pneumatic tubes are mounted on the wall: a green tube labeled "Yes" and a red tube labeled "No" (\autoref{fig:teaser}b). To grant permission, the player inserts the printed notice into the green tube; to refuse, they place it in the red tube.
    \item[Design 3] When teleported into the room, the player is positioned in front of a whiteboard displaying information about the data that will be collected in the room (\autoref{fig:wingspan_room}a). 
    The telemetry data, e.g., height and wingspan, is visually represented through a stick figure drawing.
    To accept or deny data collection, players interact with two erasers: a green eraser for acceptance and a red eraser for denial.
    If the player uses the red eraser to remove the information and, thus, denies data collection, an alternative way of solving the puzzle is triggered. 
    Instead of physically performing the actions with their body, the player must draw the required poses on the whiteboard to uncover the password.
\end{description}

\subsubsection{Challenges}
We identified five themes that addressed challenges and points to consider when creating privacy mechanisms for VR environments: \emph{Problems, Risks, Responsibility, Trade-offs,} and \emph{Questions}.

\begin{description}
    \item[Problems] The experts identified two areas: user-related and designer/developer-related challenges.
    
    From the user perspective, some players might not want to deal with privacy notices (E2). In combination with being unable to progress within the game or application without making a privacy choice, users might rush through notices without fully considering them, as mentioned by E4: "I do not think that you make people really carefully read the permissions". Additionally, with VR sensors collecting various types of data, privacy mechanisms must avoid simply replicating existing issues —"I do not want another Cookie Banner" (E1).
    Even with privacy mechanisms in place, user trust remains a challenge. Some users may doubt whether the software actually preserves privacy, preferring to rely on the OS to control sensor access: "I would only trust the OS to allow certain sensors to be accessed by the game" (E1).
    
    From the developer’s perspective, the experts acknowledged the high burden on developers (E1). They must balance user needs and preferences with company demands (E3). 
    This can mean finding a solution that maintains immersion during privacy interactions without compromising the gameplay experience.
    However, the privacy mechanism should not be too similar to the game mechanics, as this might spoil the experience ("have something similar, but not in the context of the game, not to spoil it" (E4)).
    
    When considering improvements or alternative designs, the experts mentioned they found it "hard to tell how [the mechanism] would actually look in-game" (E2), making the design process harder. This also applies to the design of the alternative game elements, which would need individual solutions depending on the game mechanics (E3) and missing sensor data in privacy-preserving scenarios.

    \item[Risks] The experts identified several risks associated with privacy mechanisms in VR environments. While some, like manipulating users by "hav[ing] hidden information in the text, like fine print" (E3) or purposefully designing mechanisms in a non-understandable way (E3), are common across digital platforms, others are unique to VR due to their immersive nature. Seamlessly integrated privacy mechanisms might blur the line between VR and real-world implications of the data collection — "In particular in an immersive game it might not be clear if this is in-game or out of the game" (E2). E2 also argued that to avoid this, it might be necessary to "break the fourth wall" and thus the immersion. This goes together with the experts' concern that 
    over-gamification could cause users not to take the privacy decision seriously (E1). Additionally, the experts noted that even alternative game elements—designed to avoid certain types of data collection—"might still reveal other data" (E2), which would need to be covered by privacy mechanisms. Finally, they raised concerns about how these mechanisms would adapt to multiplayer scenarios, e.g., for social VR (E4).
    
    \item[Responsibility] A reoccurring discussion point was the question of responsibility. With multiple stakeholders involved, the experts discussed that the responsibility could lie with the app developers or device manufacturers. One perspective was that the OS should be the trusted party handling privacy mechanism: "I would only want certain privacy and sensor level based decisions to be asked once by the OS" (E1). Another perspective involved legal requirements (E1), which ideally would cover all the necessary aspects (but would still leave the implementation with one of the other parties).

    \item[Trade-offs] The experts highlighted several trade-offs when designing privacy mechanisms, as not all aspects align. One key challenge is balancing complexity and precision with comprehension and understandability: "How can I have a nice trade-off between complexity and comprehension for the user?" (E3). This trade-off also affects implementation: "If you want to make it granular for sensors, it is the easiest to implement, but the hardest to understand" (E2). Another concern was the point at which the user is overwhelmed or even frightened by the information provided which "is kind of in contrast to the transparency the company should give" (E3).
    The experts also discussed design trade-offs, such as whether to use global settings — to avoid overwhelming users with too
    many decisions — or local settings —  to allow for granular privacy controls closely positioned to the corresponding game elements (E2). Another key balance is between making privacy interactions engaging and fun while ensuring they remain serious and effective as privacy mechanisms (E1).

    Lastly, how design decisions are made depends on the trade-off between user and company interests, which may not always align: "I guess a lot of recommendations are quite obvious, but we also know that they will not happen (...) because of the differences between what [the] company wants and what [the] user wants" (E3).
\end{description}

\subsubsection{Implications}
\label{focusgroup_recommendations}
The privacy experts discussed the three designs using the SWOT matrix as a starting point, identifying both strengths and weaknesses.
Based on these evaluations and their knowledge about privacy best practices, they identified a set of design implications.
We identified six code groups within this theme, covering different aspects targeted by the design implications, namely \emph{Getting User Consent, Information, Timing, Decision Modification, XR Context,} and \emph{Physical Control}.

\begin{description}
    \item[Getting User Consent] is the main purpose of the privacy mechanism. The experts mentioned important aspects
    of achieving this in a privacy-preserving way. Requiring the user to provide explicit consent was noted as a positive aspect of the designs several times: "I like that [Design 1] forces people to interact. It makes the decision explicit" (E2). This closely relates to the mentioned privacy-by-design approach, including things such as "provid[ing] a privacy-preserving default" (E1), where permissions are declined unless the user actively opts in.
    Another key aspect is ensuring that permissions are tied to a specific purpose, with clear details provided in the privacy mechanism (E2). This can be reinforced by closely tying the privacy decision to the corresponding interaction through timing and placement (E2). Additionally, the privacy mechanism should allow for granular decisions instead of an all-or-nothing approach and reflect this in the design in a clear way (E4).
    
    \item[Information] To provide informed consent, a clear understanding on the user's side is needed.
    As already mentioned, the notice needs to contain information about the purpose of the data collection ("The main question is the purpose, so what is done with [the data]? " (E2)) and "make it clear what the consequences are for the participants if they allow or deny specific permissions" (E4).
    When designing immersive solutions, special attention is needed to avoid having immersive but unclear designs: If symbols are used, their meaning has to be clear and unambiguous (E4) to avoid confusion or even deceptive designs. In VR environments, where the collected data may vary depending on the user's position, it is "important to let the player always be aware of the state of tracking" (E3). This could be achieved by (visually) communicating an ongoing recording or notifying the user when a sensor is activated (E1).
    
    \item[Timing] The experts generally recommended placing the privacy mechanism at the start of the game (or room), though alternative approaches could be useful depending on the context.
    Using the timing to communicate which interaction belongs to which permission was mentioned as an effective way to leverage the VR environment to enhance the design of the privacy mechanisms: "One opportunity would be to tie this interaction not to a room, but to the interaction where the data is measured" (E2). 
    
    \item[Decision Modification] The privacy experts strongly recommended allowing users to review, revoke, and modify their privacy decisions after their initial choice. This includes providing access to previously collected data and the ability to delete it if consent is withdrawn: "You can change your mind, deny your permission, and then the recorded data will be deleted for that user" (E4).
    It is important to note that some of these recommendations align with existing legal requirements in some regions.
    
    \item[VR Context] Designing privacy mechanisms for VR presents unique challenges compared to traditional 2D environments. Experts proposed several recommendations to address these challenges while leveraging VR's immersive nature. One major aspect is gamification, which, while holds the risk of distracting the user from the original purpose of informing and providing control, also has the potential to "keep it interactive and (...) make it more interesting" (E4), provided that appropriate design elements are used. Using gamification can allow the privacy interaction to become engaging and thus increase the users' awareness and attention towards privacy.
    
    To integrate the privacy decision into the VR application, using it as a tutorial for interaction methods was suggested, for example, by E2: "It reminds me of tutorials (...), this is how you do it, and by the way, this is what you are revealing by doing it". To avoid spoiling too much of the game, the interaction context should be similar but still on an abstract level (E4). Other examples included short animations or interactive videos that clarify the meaning of symbols and icons (E4) and a "physical representation of your privacy decision" (E2) to make it more tangible for the user.
    While interactive elements within the privacy mechanism can be appealing, it is important that the "approach works well for different numbers of participants and different numbers of icons" (E4) and scales well with an increasing number of permissions.
    Also, the more closely related the collected data is to the game mechanics, the more important it is to provide alternative game elements alongside the privacy mechanism to "make the app or the game usable, even without giving a single permission" (E3).
    
    \item[Physical Control] Experts noted that existing privacy mechanisms often depend on trust that the application genuinely deactivates denied sensors (E3). Implementing physical controls—such as unplugging sensors—removes this reliance, reassuring users that their choices are upheld. Alternatively, obfuscation techniques via middleware (e.g., using normalized data values (E2)) also protect privacy but shift trust to the middleware. These strategies enable users to continue using the application without disclosing all data.
\end{description}

\subsubsection{Improvements}
To make the recommendations from section \ref{focusgroup_recommendations} more concrete and address the critique points mentioned about the specific designs, we also captured the privacy experts' suggestions for improvement. This section will briefly describe design improvements using specific examples. The improvements can, of course, be applied to more than one of the designs.

Based on the discussion about the first design, three potential improvements were brought up: Instead of asking for the privacy decision at the door before the user enters the room, the experts suggested to \emph{move the choice of stickers closer to the corresponding game interaction} so it would be clearer what would happen in the room, how the data would be collected, and how it would be used in the game context ("then it is more clear why it is used and what for" (E2)). Next, the experts discussed that the user should \emph{be able to change their decision later}, which would mean that they can switch out the stickers anytime, "to have the opportunity also to change your mind after entering the room" (E4). The last improvement suggested for the first design would allow the user to \emph{make granular decisions} by choosing the individual stickers instead of taking the entire sheet from the door (E2). 

When the experts discussed design number two, three ways to improve the interaction emerged. To implement a \emph{default setting to less data collection}, the experts suggested an approach where only the green tube would be needed. Like this, only if the user wanted to give \emph{explicit consent} to the data collection they had to do something (send the permission through the green tube) otherwise no action would be needed ("I just give active permission to the things I want to use. But I do not need to decline permissions" (E3)). As a third improvement, the experts suggested that similar to the first design, the \emph{use of icons} could help to make the interaction more engaging (E2).

For the third design, the experts did not mention specific improvements, but their critique points also provided insight into how the design might be improved. Although they thought of the eraser as a fitting interaction mechanic to delete permission or collected data, using the green eraser to give consent seemed counterintuitive (E1), so the experts proposed choosing a more suitable interaction mechanism. To allow the user to review their previously made decision, they also suggested "take a picture [of the flip chart] or just document it somewhere" (E4) for future access.

\subsubsection{Overlap with Design Session Results}
We identified several common themes in the viewpoints of designers and privacy experts. Both groups emphasized clarifying \textit{responsibility} for privacy decisions and recognizing the \textit{trade-offs} intrinsic to the design process. They also converged on \emph{recommendations} for increasing transparency, ensuring users can revise their privacy choices at any time, and motivating users to engage more thoughtfully with the privacy mechanism.

\subsection{Limitations}
We outline two limitations related to \textit{(1) our participant sample} and \textit{(2) the generalizability of our study context}.

 Our concept brainstorming and sketching sessions involved student game designers and developers with at least one semester of relevant coursework. Although their professional experience was limited, many challenges they described mirror those documented in prior work with more experienced practitioners~\cite{adams2018ethics}. To mitigate potential gaps in expertise, we conducted a complementary focus group with four usable privacy experts, including one with XR experience. Their consensus on our proposed designs increases the strength of the implications.

We anchored our exploration in a \textit{VR escape room game} to create a concrete, engaging environment for evaluating privacy interface concepts. While this approach yielded insights and novel concepts—such as the privacy sticker book—certain game-specific elements may not directly extend to other VR domains (e.g., professional training, education). Nonetheless, our results suggest that abstracting core ideas, such as visual indicators of sensor usage or physical interactions for consent, can inform designs beyond gaming. We do not assert the universal applicability of our scenario-based elicitation method across all XR use cases, particularly those beyond head-worn VR. However, our work demonstrates how existing notions (e.g., camera indicator lights) can be adapted to VR’s unique affordances, such as visually trailing the user’s avatar. To further enhance generalizability, future work could investigate a wider range of scenarios and introduce priming tasks that more broadly capture different design objectives, ensuring applicability across diverse XR environments.

\section{Discussion and Implications}
In this study, we adopted a hands-on approach to designing privacy interfaces for VR, to reveal and address the unique privacy challenges posed by immersive technologies. We critically examine the implications of our findings and outline directions for future research. Some implications align with established best practices for privacy interfaces in other contexts. The first group (Implications 1–3) synthesizes overlapping insights from the design sessions and focus groups, tailoring them to immersive VR settings. The second group (Implications 4–7) highlights considerations that are either VR-specific or introduce novel dynamics, reflecting the distinctive affordances and challenges of immersive environments.

\subsection{Privacy by Design and Data Minimization in VR: Ensuring User-Centric Data Collection}

Our findings underscore the importance of embedding privacy by design principles when creating VR environments, emphasizing minimal data collection and strong user control. Expert recommendations endorse \textit{privacy-preserving defaults}, which limit data gathering to what is essential for core functionality, with any additional collection necessitating \textit{explicit user consent}. They also advocate \textit{purpose-tied consent mechanisms}, where consent applies only to the specific features or functionalities in question, rather than broad or generalized permissions. These insights align with earlier work~\cite{cavoukian2009privacy} and reflect requirements under privacy laws such as the EU’s GDPR\footnote{GDPR: \url{https://gdpr-info.eu/}}, highlighting their relevance to VR privacy challenges.

Feedback from the design sessions revealed that many participants were unfamiliar with these principles or had never applied them in their designs. To foster broader adoption, future work could explore automated tools that guide developers in implementing privacy-preserving methods during VR development, akin to \textit{Reframe}~\cite{rajaram2023reframe}, which supports privacy and security threat modeling for AR. Further research is also needed to assess user comprehension and interaction with privacy-preserving VR interfaces, and to determine how these methods influence overall user experience and engagement.

\summaryBox{1: Use Explicit and Purpose-Tied Consent Mechanisms}{Ensure that each data collection request is clearly tied to a specific purpose, with a detailed explanation of the data being collected and why. Employ a “privacy by default” approach, with data collection set to the minimum necessary by default unless explicitly consented to by the user.}

A persistent issue in our study was the reliance on user trust that privacy choices are respected in practice. For instance, users have no direct guarantee that denying sensor permission truly halts data collection—an especially difficult problem in VR, where sensor arrays are often sophisticated and continuously active. Experts highlighted \textit{physical controls}—such as unplugging sensors or covering lenses—to provide visible and tangible evidence that a user’s choice is upheld. While such approaches reflect prior work on tangible privacy mechanisms~\cite{mehtaprivacycare2021, delgado2022prikey}, they are not always practical in complex VR systems where sensors are deeply integrated.

This underscores a larger gap in VR privacy design: the need for \textit{verifiable} enforcement of user preferences. Participants frequently pointed out the limited ways for users to confirm whether their privacy settings were actually being respected. Future research could address this gap by developing \textit{real-time audit tools} (e.g., interactive dashboards displaying active sensors and data flows) or by creating \textit{verifiable logs} that certify how and when data was collected. \textit{Middleware-based data anonymization}, which obfuscates sensitive information before it reaches the application, could add another layer of privacy protection. Likewise, \textit{centralized privacy management at the OS level} can unify and enforce privacy settings across multiple XR applications.

By integrating real-time monitoring, verifiable logging, and anonymization techniques, VR systems can enhance both trust and accountability, ensuring that privacy choices become more than just policy statements—they become actions users can verify.

\summaryBox{2: Include Verifiable Privacy Controls}{Provide users with tools to verify that their privacy settings are being adhered to, such as real-time audit tools and verifiable data logs. Where possible, offer physical or technical solutions, such as sensor covers or middleware that anonymizes data, to enhance user trust.}

\subsection{Granular Control and Transparency: Balancing Detail with Usability.} 
Focus group discussions underscored the value of \textit{granular control} over privacy settings, allowing users to specify exactly which data to share and to adjust their choices over time. This approach gives users comprehensive autonomy, including straightforward methods to revoke consent or delete previously collected data. However, in fast-paced VR environments with fragmented attention, such flexibility may cause confusion or disengagement if the interface becomes too complex.

To mitigate these issues, privacy interfaces should balance detailed control and ease of use. Proposed strategies include \textit{details on demand}—revealing additional information only when necessary—and \textit{do not ask again} options to minimize repetitive prompts. \textit{Adaptive privacy interfaces} can further alleviate decision fatigue by tailoring consent requests to user behavior and preferences, following approaches similar to mobile app recommendations~\cite{liu2016}. In addition, \textit{continuous feedback}—for instance, ambient cues indicating active data collection—can maintain awareness without overwhelming the user. Such cues are particularly crucial in VR, where full immersion may obscure real-time changes in collected data~\cite{nair2023exploring}.

\summaryBox{3: Provide Clear, Granular, and Modifiable Consent Options}{Offer detailed control over data permissions, allowing users to make informed choices while prioritizing critical settings to avoid overwhelming them. Allow users to revisit and adjust their privacy choices at any time, including options to revoke consent and delete previously collected data.}

\summaryBox{4: Enhance Transparency with Continuous Feedback}{Provide real-time, context-sensitive feedback on the status of sensors and data collection within the VR environment to maintain user awareness and build trust.}

\subsection{Gamification of Privacy Decisions: Enhancing Engagement or Oversimplifying Consent?}
Designers and developers proposed using gamification elements, such as stickers and pneumatic mailing tubes, to incentivize users to engage more actively in privacy decisions. 
These interactive elements hold particular value in VR contexts, where traditional long text-based interfaces can suffer from poor readability~\cite{Kojic2020}. 
Previous studies have similarly explored gamified visualizations of collected data to improve the understandability of privacy policies~\cite{lim2022mine, morrison2014improving}. However, these approaches primarily focus on raising awareness rather than enabling control. 

While integrating privacy decisions into game mechanics can make privacy interactions more engaging~\cite{kitkowska2020}, it poses the risk that users may not fully comprehend the implications of their choices, treating them as just another task to complete in the game rather than an important decision about their personal data.
To mitigate this risk, it is important to \textit{gamify privacy decisions thoughtfully}, ensuring that gamified interactions do not trivialize privacy choices or distract from critical information about data collection. 
Designers should carefully balance the engaging aspects of gamification with the need for clear communication of privacy risks and implications.


\summaryBox{5: Gamify Privacy Decisions Thoughtfully}{Incorporate gamification elements to engage users in making privacy decisions and increase usability in VR; ensure that gamified interactions do not trivialize privacy choices or distract from critical information about data collection.}

\subsection{Accessibility in VR: Bridging the Gap for All Users}
Interface accessibility emerged as a theme in the design process, particularly as designers reflected on the need to accommodate users with diverse abilities through multimodal feedback for data collection awareness, such as combining visual, auditory, or haptic
feedback.
Barriers to accessibility can lead to misinterpretation of privacy controls, unintended data sharing, or complete disengagement from privacy settings, leaving users vulnerable.
Ensuring that VR interfaces are accessible to users with different levels of technical expertise, and physical and cognitive abilities helps make these environments more inclusive. 
While accessibility is not a new requirement, it takes on new significance in VR due to its added interaction modalities. Designers must carefully balance the expanded design possibilities in VR with the need for clear, intuitive, and universally accessible privacy interfaces.

To maintain accessibility, VR privacy interfaces should prioritize simplicity in design, providing clear and intuitive controls that are easy to understand and interact with, even for novice users. 
Given the relatively recent availability of consumer-grade VR technology and its growing adoption by novice users, additional attention should be directed towards explaining how data is collected and used, as many users may not be as familiar with these processes compared to 2D environments.
Co-designing and testing these interfaces with diverse user groups is essential to ensure that accessibility is also maintained as the technology evolves.

\summaryBox{6: Address Accessibility and User Understanding}{Design the VR data collection awareness and control interface for the needs of diverse user groups to ensure that they are accessible to users with varying levels of technical knowledge and abilities.}

\subsection{Scalability in VR: Managing Growth and Complexity}

Although scalability of privacy permissions is not a new challenge~\cite{prange2024soups}, VR environments introduce additional complexities. Designers must handle an expanding range of sensors and permissions without compromising usability. Experts also emphasized the importance of adapting control mechanisms to multi-agent scenarios like social VR, which extend privacy concerns to bystanders and other participants. As VR applications grow in scale and sophistication, privacy controls must keep pace, managing increasing data flows and multi-user dynamics. Yet, scaling can conflict with simplicity, as layered controls may overwhelm users and hinder comprehension. Consequently, VR privacy interfaces must be engineered to handle both proliferation of sensors and broader user groups while retaining user-friendly features.

\summaryBox{7: Design for Scalability}{Ensure that the control mechanism scales effectively for an increasing number of sensors and permissions. 
Consider the long-term implications of privacy controls in multi-user environments, such as shared XR spaces or social VR applications.}

\section{Future Work}

Our findings indicate several promising avenues for further investigation beyond the scope of this paper. First, it would be valuable to validate the proposed design concepts and implications across a broader spectrum of VR contexts—such as educational, professional, and social applications—beyond gaming. User evaluations in these settings could serve as a basis for refining and extending the implications into more targeted design guidelines.

Second, future work could explore automated tools and frameworks guiding developers in incorporating privacy-preserving principles into VR apps, potentially leveraging established practices from other domains.

Third, research could investigate how different levels of gamification affect user engagement and comprehension in VR privacy interfaces. While introducing gamified elements may enhance interactivity, they also risk oversimplifying critical privacy decisions. Balancing user enjoyment with thorough understanding is essential for maintaining the seriousness of privacy choices.

Finally, longitudinal studies might offer insights into how these interfaces influence user behavior and trust over time, especially as XR technologies become more pervasive. Such studies could shed light on how sustained exposure shapes privacy practices and overall user trust in XR systems.



\section{Conclusion}

This paper examined how to redesign data collection awareness and control interfaces for VR, pinpointing the challenges and opportunities inherent in immersive environments. Through concept brainstorming and sketching sessions with novice game developers and designers, followed by a focus group with privacy experts, we identified key issues such as balancing user engagement with privacy awareness, handling complex privacy details without overwhelming users, and ensuring trust, transparency, and regulatory compliance.

We proposed design implications that include explicit, purpose-tied consent mechanisms, continuous feedback on data collection status, and alternative data-minimizing game tasks. We also advocate careful gamification to enhance user engagement without trivializing privacy decisions, as well as granular, verifiable controls to foster trust and give users effective data management options. Future work could investigate these design concepts in broader XR contexts, explore different gamification approaches on user engagement and comprehension, and develop tools to support privacy-preserving design. Overall, our findings contribute to the growing body of research on usable privacy in XR, offering actionable guidance for creating more privacy-aware, inclusive, and user-friendly VR experiences.

\begin{acks}
The authors would like to thank all participants in the study.
\end{acks}


\bibliographystyle{ACM-Reference-Format}
\bibliography{bibliography}


\begin{thebibliography}{49}


\ifx \showCODEN    \undefined \def \showCODEN     #1{\unskip}     \fi
\ifx \showDOI      \undefined \def \showDOI       #1{#1}\fi
\ifx \showISBNx    \undefined \def \showISBNx     #1{\unskip}     \fi
\ifx \showISBNxiii \undefined \def \showISBNxiii  #1{\unskip}     \fi
\ifx \showISSN     \undefined \def \showISSN      #1{\unskip}     \fi
\ifx \showLCCN     \undefined \def \showLCCN      #1{\unskip}     \fi
\ifx \shownote     \undefined \def \shownote      #1{#1}          \fi
\ifx \showarticletitle \undefined \def \showarticletitle #1{#1}   \fi
\ifx \showURL      \undefined \def \showURL       {\relax}        \fi
\providecommand\bibfield[2]{#2}
\providecommand\bibinfo[2]{#2}
\providecommand\natexlab[1]{#1}
\providecommand\showeprint[2][]{arXiv:#2}

\bibitem[Abraham et~al\mbox{.}(2024)]%
        {abraham2024}
\bibfield{author}{\bibinfo{person}{Melvin Abraham}, \bibinfo{person}{Mark Mcgill}, {and} \bibinfo{person}{Mohamed Khamis}.} \bibinfo{year}{2024}\natexlab{}.
\newblock \showarticletitle{What You Experience is What We Collect: User Experience Based Fine-Grained Permissions for Everyday Augmented Reality}. In \bibinfo{booktitle}{\emph{Proceedings of the CHI Conference on Human Factors in Computing Systems}} (Honolulu, HI, USA) \emph{(\bibinfo{series}{CHI '24})}. \bibinfo{publisher}{Association for Computing Machinery}, \bibinfo{address}{New York, NY, USA}, Article \bibinfo{articleno}{772}, \bibinfo{numpages}{24}~pages.
\newblock
\showISBNx{9798400703300}
\urldef\tempurl%
\url{https://doi.org/10.1145/3613904.3642668}
\showDOI{\tempurl}


\bibitem[Abraham et~al\mbox{.}(2022)]%
        {abraham2022implications}
\bibfield{author}{\bibinfo{person}{Melvin Abraham}, \bibinfo{person}{Pejman Saeghe}, \bibinfo{person}{Mark Mcgill}, {and} \bibinfo{person}{Mohamed Khamis}.} \bibinfo{year}{2022}\natexlab{}.
\newblock \showarticletitle{Implications of XR on Privacy, Security and Behaviour: Insights from Experts}. In \bibinfo{booktitle}{\emph{Nordic Human-Computer Interaction Conference}} (Aarhus, Denmark) \emph{(\bibinfo{series}{NordiCHI '22})}. \bibinfo{publisher}{Association for Computing Machinery}, \bibinfo{address}{New York, NY, USA}, Article \bibinfo{articleno}{30}, \bibinfo{numpages}{12}~pages.
\newblock
\showISBNx{9781450396998}
\urldef\tempurl%
\url{https://doi.org/10.1145/3546155.3546691}
\showDOI{\tempurl}


\bibitem[Adams et~al\mbox{.}(2018)]%
        {adams2018ethics}
\bibfield{author}{\bibinfo{person}{Devon Adams}, \bibinfo{person}{Alseny Bah}, \bibinfo{person}{Catherine Barwulor}, \bibinfo{person}{Nureli Musaby}, \bibinfo{person}{Kadeem Pitkin}, {and} \bibinfo{person}{Elissa~M. Redmiles}.} \bibinfo{year}{2018}\natexlab{}.
\newblock \showarticletitle{Ethics Emerging: the Story of Privacy and Security Perceptions in Virtual Reality}. In \bibinfo{booktitle}{\emph{Fourteenth Symposium on Usable Privacy and Security (SOUPS 2018)}}. \bibinfo{publisher}{USENIX Association}, \bibinfo{address}{Baltimore, MD}, \bibinfo{pages}{427--442}.
\newblock
\showISBNx{978-1-939133-10-6}
\urldef\tempurl%
\url{https://www.usenix.org/conference/soups2018/presentation/adams}
\showURL{%
\tempurl}


\bibitem[Cavoukian(2009)]%
        {cavoukian2009privacy}
\bibfield{author}{\bibinfo{person}{Ann Cavoukian}.} \bibinfo{year}{2009}\natexlab{}.
\newblock \showarticletitle{Privacy by design: The 7 foundational principles}.
\newblock \bibinfo{journal}{\emph{Information and privacy commissioner of Ontario, Canada}}  \bibinfo{volume}{5} (\bibinfo{year}{2009}), \bibinfo{pages}{2009}.
\newblock


\bibitem[De~Guzman et~al\mbox{.}(2019)]%
        {guzman2019}
\bibfield{author}{\bibinfo{person}{Jaybie~A. De~Guzman}, \bibinfo{person}{Kanchana Thilakarathna}, {and} \bibinfo{person}{Aruna Seneviratne}.} \bibinfo{year}{2019}\natexlab{}.
\newblock \showarticletitle{Security and Privacy Approaches in Mixed Reality: A Literature Survey}.
\newblock \bibinfo{journal}{\emph{ACM Comput. Surv.}} \bibinfo{volume}{52}, \bibinfo{number}{6}, Article \bibinfo{articleno}{110} (\bibinfo{date}{oct} \bibinfo{year}{2019}), \bibinfo{numpages}{37}~pages.
\newblock
\showISSN{0360-0300}
\urldef\tempurl%
\url{https://doi.org/10.1145/3359626}
\showDOI{\tempurl}


\bibitem[Delgado~Rodriguez et~al\mbox{.}(2022)]%
        {delgado2022prikey}
\bibfield{author}{\bibinfo{person}{Sarah Delgado~Rodriguez}, \bibinfo{person}{Sarah Prange}, \bibinfo{person}{Christina Vergara~Ossenberg}, \bibinfo{person}{Markus Henkel}, \bibinfo{person}{Florian Alt}, {and} \bibinfo{person}{Karola Marky}.} \bibinfo{year}{2022}\natexlab{}.
\newblock \showarticletitle{PriKey – Investigating Tangible Privacy Control for Smart Home Inhabitants and Visitors}. In \bibinfo{booktitle}{\emph{Nordic Human-Computer Interaction Conference}} (Aarhus, Denmark) \emph{(\bibinfo{series}{NordiCHI '22})}. \bibinfo{publisher}{Association for Computing Machinery}, \bibinfo{address}{New York, NY, USA}, Article \bibinfo{articleno}{74}, \bibinfo{numpages}{13}~pages.
\newblock
\showISBNx{9781450396998}
\urldef\tempurl%
\url{https://doi.org/10.1145/3546155.3546640}
\showDOI{\tempurl}


\bibitem[Farke et~al\mbox{.}(2021)]%
        {farke2021}
\bibfield{author}{\bibinfo{person}{Florian~M. Farke}, \bibinfo{person}{David~G. Balash}, \bibinfo{person}{Maximilian Golla}, \bibinfo{person}{Markus D{\"u}rmuth}, {and} \bibinfo{person}{Adam~J. Aviv}.} \bibinfo{year}{2021}\natexlab{}.
\newblock \showarticletitle{Are Privacy Dashboards Good for End Users? Evaluating User Perceptions and Reactions to Google{\textquoteright}s My Activity}. In \bibinfo{booktitle}{\emph{30th USENIX Security Symposium (USENIX Security 21)}}. \bibinfo{publisher}{USENIX Association}, \bibinfo{pages}{483--500}.
\newblock
\showISBNx{978-1-939133-24-3}
\urldef\tempurl%
\url{https://www.usenix.org/conference/usenixsecurity21/presentation/farke}
\showURL{%
\tempurl}


\bibitem[Fassl et~al\mbox{.}(2021)]%
        {fassl2021exploring}
\bibfield{author}{\bibinfo{person}{Matthias Fassl}, \bibinfo{person}{Lea~Theresa Gr\"{o}ber}, {and} \bibinfo{person}{Katharina Krombholz}.} \bibinfo{year}{2021}\natexlab{}.
\newblock \showarticletitle{Exploring User-Centered Security Design for Usable Authentication Ceremonies}. In \bibinfo{booktitle}{\emph{Proceedings of the 2021 CHI Conference on Human Factors in Computing Systems}} (Yokohama, Japan) \emph{(\bibinfo{series}{CHI '21})}. \bibinfo{publisher}{Association for Computing Machinery}, \bibinfo{address}{New York, NY, USA}, Article \bibinfo{articleno}{694}, \bibinfo{numpages}{15}~pages.
\newblock
\showISBNx{9781450380966}
\urldef\tempurl%
\url{https://doi.org/10.1145/3411764.3445164}
\showDOI{\tempurl}


\bibitem[Feng et~al\mbox{.}(2021)]%
        {feng2021}
\bibfield{author}{\bibinfo{person}{Yuanyuan Feng}, \bibinfo{person}{Yaxing Yao}, {and} \bibinfo{person}{Norman Sadeh}.} \bibinfo{year}{2021}\natexlab{}.
\newblock \showarticletitle{A Design Space for Privacy Choices: Towards Meaningful Privacy Control in the Internet of Things}. In \bibinfo{booktitle}{\emph{Proceedings of the 2021 CHI Conference on Human Factors in Computing Systems}} (Yokohama, Japan) \emph{(\bibinfo{series}{CHI '21})}. \bibinfo{publisher}{Association for Computing Machinery}, \bibinfo{address}{New York, NY, USA}, Article \bibinfo{articleno}{64}, \bibinfo{numpages}{16}~pages.
\newblock
\showISBNx{9781450380966}
\urldef\tempurl%
\url{https://doi.org/10.1145/3411764.3445148}
\showDOI{\tempurl}


\bibitem[Garrido et~al\mbox{.}(2024)]%
        {garrido2024sok}
\bibfield{author}{\bibinfo{person}{Gonzalo~Munilla Garrido}, \bibinfo{person}{Vivek Nair}, {and} \bibinfo{person}{Dawn Song}.} \bibinfo{year}{2024}\natexlab{}.
\newblock \showarticletitle{SoK: Data Privacy in Virtual Reality}.
\newblock \bibinfo{journal}{\emph{Proceedings on Privacy Enhancing Technologies}} (\bibinfo{year}{2024}).
\newblock


\bibitem[Gisch et~al\mbox{.}(2007)]%
        {gisch2007}
\bibfield{author}{\bibinfo{person}{Martin Gisch}, \bibinfo{person}{Alexander De~Luca}, {and} \bibinfo{person}{Markus Blanchebarbe}.} \bibinfo{year}{2007}\natexlab{}.
\newblock \showarticletitle{The privacy badge: a privacy-awareness user interface for small devices}. In \bibinfo{booktitle}{\emph{Proceedings of the 4th International Conference on Mobile Technology, Applications, and Systems and the 1st International Symposium on Computer Human Interaction in Mobile Technology}} (Singapore) \emph{(\bibinfo{series}{Mobility '07})}. \bibinfo{publisher}{Association for Computing Machinery}, \bibinfo{address}{New York, NY, USA}, \bibinfo{pages}{583–586}.
\newblock
\showISBNx{9781595938190}
\urldef\tempurl%
\url{https://doi.org/10.1145/1378063.1378159}
\showDOI{\tempurl}


\bibitem[Hadan et~al\mbox{.}(2024)]%
        {hadan2024privacy}
\bibfield{author}{\bibinfo{person}{Hilda Hadan}, \bibinfo{person}{Derrick~M. Wang}, \bibinfo{person}{Lennart~E. Nacke}, {and} \bibinfo{person}{Leah Zhang-Kennedy}.} \bibinfo{year}{2024}\natexlab{}.
\newblock \showarticletitle{Privacy in Immersive Extended Reality: Exploring User Perceptions, Concerns, and Coping Strategies}. In \bibinfo{booktitle}{\emph{Proceedings of the CHI Conference on Human Factors in Computing Systems}} (Honolulu, HI, USA) \emph{(\bibinfo{series}{CHI '24})}. \bibinfo{publisher}{Association for Computing Machinery}, \bibinfo{address}{New York, NY, USA}, Article \bibinfo{articleno}{784}, \bibinfo{numpages}{24}~pages.
\newblock
\showISBNx{9798400703300}
\urldef\tempurl%
\url{https://doi.org/10.1145/3613904.3642104}
\showDOI{\tempurl}


\bibitem[Houzangbe et~al\mbox{.}(2022)]%
        {houzangbe2022}
\bibfield{author}{\bibinfo{person}{Samory Houzangbe}, \bibinfo{person}{Dimitri Masson}, \bibinfo{person}{Sylvain Fleury}, \bibinfo{person}{David~Antonio Gómez~Jáuregui}, \bibinfo{person}{Jeremy Legardeur}, \bibinfo{person}{Simon Richir}, {and} \bibinfo{person}{Nadine Couture}.} \bibinfo{year}{2022}\natexlab{}.
\newblock \showarticletitle{Is virtual reality the solution? A comparison between 3D and 2D creative sketching tools in the early design process}.
\newblock \bibinfo{journal}{\emph{Frontiers in Virtual Reality}}  \bibinfo{volume}{3} (\bibinfo{year}{2022}).
\newblock
\showISSN{2673-4192}
\urldef\tempurl%
\url{https://doi.org/10.3389/frvir.2022.958223}
\showDOI{\tempurl}


\bibitem[Kelley et~al\mbox{.}(2009)]%
        {kelley2009}
\bibfield{author}{\bibinfo{person}{Patrick~Gage Kelley}, \bibinfo{person}{Joanna Bresee}, \bibinfo{person}{Lorrie~Faith Cranor}, {and} \bibinfo{person}{Robert~W. Reeder}.} \bibinfo{year}{2009}\natexlab{}.
\newblock \showarticletitle{A "nutrition label" for privacy}. In \bibinfo{booktitle}{\emph{Proceedings of the 5th Symposium on Usable Privacy and Security}} (Mountain View, California, USA) \emph{(\bibinfo{series}{SOUPS '09})}. \bibinfo{publisher}{Association for Computing Machinery}, \bibinfo{address}{New York, NY, USA}, Article \bibinfo{articleno}{4}, \bibinfo{numpages}{12}~pages.
\newblock
\showISBNx{9781605587363}
\urldef\tempurl%
\url{https://doi.org/10.1145/1572532.1572538}
\showDOI{\tempurl}


\bibitem[King et~al\mbox{.}(2011)]%
        {king2011}
\bibfield{author}{\bibinfo{person}{Jennifer King}, \bibinfo{person}{Airi Lampinen}, {and} \bibinfo{person}{Alex Smolen}.} \bibinfo{year}{2011}\natexlab{}.
\newblock \showarticletitle{Privacy: is there an app for that?}. In \bibinfo{booktitle}{\emph{Proceedings of the Seventh Symposium on Usable Privacy and Security}} (Pittsburgh, Pennsylvania) \emph{(\bibinfo{series}{SOUPS '11})}. \bibinfo{publisher}{Association for Computing Machinery}, \bibinfo{address}{New York, NY, USA}, Article \bibinfo{articleno}{12}, \bibinfo{numpages}{20}~pages.
\newblock
\showISBNx{9781450309110}
\urldef\tempurl%
\url{https://doi.org/10.1145/2078827.2078843}
\showDOI{\tempurl}


\bibitem[Kitkowska et~al\mbox{.}(2020)]%
        {kitkowska2020}
\bibfield{author}{\bibinfo{person}{Agnieszka Kitkowska}, \bibinfo{person}{Mark Warner}, \bibinfo{person}{Yefim Shulman}, \bibinfo{person}{Erik W{\"a}stlund}, {and} \bibinfo{person}{Leonardo~A. Martucci}.} \bibinfo{year}{2020}\natexlab{}.
\newblock \showarticletitle{Enhancing Privacy through the Visual Design of Privacy Notices: Exploring the Interplay of Curiosity, Control and Affect}. In \bibinfo{booktitle}{\emph{Sixteenth Symposium on Usable Privacy and Security (SOUPS 2020)}}. \bibinfo{publisher}{USENIX Association}, \bibinfo{pages}{437--456}.
\newblock
\showISBNx{978-1-939133-16-8}
\urldef\tempurl%
\url{https://www.usenix.org/conference/soups2020/presentation/kitkowska}
\showURL{%
\tempurl}


\bibitem[Kojić et~al\mbox{.}(2020)]%
        {Kojic2020}
\bibfield{author}{\bibinfo{person}{Tanja Kojić}, \bibinfo{person}{Danish Ali}, \bibinfo{person}{Robert Greinacher}, \bibinfo{person}{Sebastian Möller}, {and} \bibinfo{person}{Jan-Niklas Voigt-Antons}.} \bibinfo{year}{2020}\natexlab{}.
\newblock \showarticletitle{User Experience of Reading in Virtual Reality — Finding Values for Text Distance, Size and Contrast}. In \bibinfo{booktitle}{\emph{2020 Twelfth International Conference on Quality of Multimedia Experience (QoMEX)}}. \bibinfo{pages}{1--6}.
\newblock
\urldef\tempurl%
\url{https://doi.org/10.1109/QoMEX48832.2020.9123091}
\showDOI{\tempurl}


\bibitem[Kollnig et~al\mbox{.}(2022)]%
        {kolling2022}
\bibfield{author}{\bibinfo{person}{Konrad Kollnig}, \bibinfo{person}{Anastasia Shuba}, \bibinfo{person}{Max Van~Kleek}, \bibinfo{person}{Reuben Binns}, {and} \bibinfo{person}{Nigel Shadbolt}.} \bibinfo{year}{2022}\natexlab{}.
\newblock \showarticletitle{Goodbye Tracking? Impact of iOS App Tracking Transparency and Privacy Labels}. In \bibinfo{booktitle}{\emph{Proceedings of the 2022 ACM Conference on Fairness, Accountability, and Transparency}} (Seoul, Republic of Korea) \emph{(\bibinfo{series}{FAccT '22})}. \bibinfo{publisher}{Association for Computing Machinery}, \bibinfo{address}{New York, NY, USA}, \bibinfo{pages}{508–520}.
\newblock
\showISBNx{9781450393522}
\urldef\tempurl%
\url{https://doi.org/10.1145/3531146.3533116}
\showDOI{\tempurl}


\bibitem[Kraus et~al\mbox{.}(2014)]%
        {kraus2014}
\bibfield{author}{\bibinfo{person}{Lydia Kraus}, \bibinfo{person}{Ina Wechsung}, {and} \bibinfo{person}{Sebastian Möller}.} \bibinfo{year}{2014}\natexlab{}.
\newblock \showarticletitle{Using Statistical Information to Communicate Android Permission Risks to Users}. In \bibinfo{booktitle}{\emph{2014 Workshop on Socio-Technical Aspects in Security and Trust}}. \bibinfo{pages}{48--55}.
\newblock
\urldef\tempurl%
\url{https://doi.org/10.1109/STAST.2014.15}
\showDOI{\tempurl}


\bibitem[Kröger et~al\mbox{.}(2023)]%
        {kroeger2023surveilling}
\bibfield{author}{\bibinfo{person}{Jacob~Leon Kröger}, \bibinfo{person}{Philip Raschke}, \bibinfo{person}{Jessica {Percy Campbell}}, {and} \bibinfo{person}{Stefan Ullrich}.} \bibinfo{year}{2023}\natexlab{}.
\newblock \showarticletitle{Surveilling the gamers: Privacy impacts of the video game industry}.
\newblock \bibinfo{journal}{\emph{Entertainment Computing}}  \bibinfo{volume}{44} (\bibinfo{year}{2023}), \bibinfo{pages}{100537}.
\newblock
\showISSN{1875-9521}
\urldef\tempurl%
\url{https://doi.org/10.1016/j.entcom.2022.100537}
\showDOI{\tempurl}


\bibitem[Lee et~al\mbox{.}(2019)]%
        {lee2019design}
\bibfield{author}{\bibinfo{person}{Jee~Hyun Lee}, \bibinfo{person}{Eun~Kyoung Yang}, {and} \bibinfo{person}{Zhong~Yuan Sun}.} \bibinfo{year}{2019}\natexlab{}.
\newblock \showarticletitle{Design Cognitive Actions Stimulating Creativity in the VR Design Environment}. In \bibinfo{booktitle}{\emph{Proceedings of the 2019 Conference on Creativity and Cognition}} (San Diego, CA, USA) \emph{(\bibinfo{series}{C\&C '19})}. \bibinfo{publisher}{Association for Computing Machinery}, \bibinfo{address}{New York, NY, USA}, \bibinfo{pages}{604–611}.
\newblock
\showISBNx{9781450359177}
\urldef\tempurl%
\url{https://doi.org/10.1145/3325480.3326575}
\showDOI{\tempurl}


\bibitem[Lim et~al\mbox{.}(2022)]%
        {lim2022mine}
\bibfield{author}{\bibinfo{person}{Junsu Lim}, \bibinfo{person}{Hyeonggeun Yun}, \bibinfo{person}{Auejin Ham}, {and} \bibinfo{person}{Sunjun Kim}.} \bibinfo{year}{2022}\natexlab{}.
\newblock \showarticletitle{Mine Yourself!: A Role-playing Privacy Tutorial in Virtual Reality Environment}. In \bibinfo{booktitle}{\emph{Extended Abstracts of the 2022 CHI Conference on Human Factors in Computing Systems}} (New Orleans, LA, USA) \emph{(\bibinfo{series}{CHI EA '22})}. \bibinfo{publisher}{Association for Computing Machinery}, \bibinfo{address}{New York, NY, USA}, Article \bibinfo{articleno}{375}, \bibinfo{numpages}{7}~pages.
\newblock
\showISBNx{9781450391566}
\urldef\tempurl%
\url{https://doi.org/10.1145/3491101.3519773}
\showDOI{\tempurl}


\bibitem[Liu et~al\mbox{.}(2016)]%
        {liu2016}
\bibfield{author}{\bibinfo{person}{Bin Liu}, \bibinfo{person}{Mads~Schaarup Andersen}, \bibinfo{person}{Florian Schaub}, \bibinfo{person}{Hazim Almuhimedi}, \bibinfo{person}{Shikun~(Aerin) Zhang}, \bibinfo{person}{Norman Sadeh}, \bibinfo{person}{Yuvraj Agarwal}, {and} \bibinfo{person}{Alessandro Acquisti}.} \bibinfo{year}{2016}\natexlab{}.
\newblock \showarticletitle{Follow My Recommendations: A Personalized Privacy Assistant for Mobile App Permissions}. In \bibinfo{booktitle}{\emph{Twelfth Symposium on Usable Privacy and Security (SOUPS 2016)}}. \bibinfo{publisher}{USENIX Association}, \bibinfo{address}{Denver, CO}, \bibinfo{pages}{27--41}.
\newblock
\showISBNx{978-1-931971-31-7}
\urldef\tempurl%
\url{https://www.usenix.org/conference/soups2016/technical-sessions/presentation/liu}
\showURL{%
\tempurl}


\bibitem[Malhotra et~al\mbox{.}(2004)]%
        {malhotra2004internet}
\bibfield{author}{\bibinfo{person}{Naresh~K Malhotra}, \bibinfo{person}{Sung~S Kim}, {and} \bibinfo{person}{James Agarwal}.} \bibinfo{year}{2004}\natexlab{}.
\newblock \showarticletitle{Internet users' information privacy concerns (IUIPC): The construct, the scale, and a causal model}.
\newblock \bibinfo{journal}{\emph{Information systems research}} \bibinfo{volume}{15}, \bibinfo{number}{4} (\bibinfo{year}{2004}), \bibinfo{pages}{336--355}.
\newblock


\bibitem[Mehta et~al\mbox{.}(2021)]%
        {mehtaprivacycare2021}
\bibfield{author}{\bibinfo{person}{Vikram Mehta}, \bibinfo{person}{Daniel Gooch}, \bibinfo{person}{Arosha Bandara}, \bibinfo{person}{Blaine Price}, {and} \bibinfo{person}{Bashar Nuseibeh}.} \bibinfo{year}{2021}\natexlab{}.
\newblock \showarticletitle{Privacy Care: A Tangible Interaction Framework for Privacy Management}.
\newblock \bibinfo{journal}{\emph{ACM Trans. Internet Technol.}} \bibinfo{volume}{21}, \bibinfo{number}{1}, Article \bibinfo{articleno}{25} (\bibinfo{date}{Feb.} \bibinfo{year}{2021}), \bibinfo{numpages}{32}~pages.
\newblock
\showISSN{1533-5399}
\urldef\tempurl%
\url{https://doi.org/10.1145/3430506}
\showDOI{\tempurl}


\bibitem[Meta(2024)]%
        {ttcLabsXR}
\bibfield{author}{\bibinfo{person}{Meta}.} \bibinfo{year}{2023 (accessed July, 2024)}\natexlab{}.
\newblock \bibinfo{title}{Data Transparency and Control in XR and the Metaverse}.
\newblock \bibinfo{howpublished}{\url{http://www.ttclabs.net/site/assets/files/11085/data_transparency_and_control_in_xr_and_the_metaverse_report.pdf}}.
\newblock


\bibitem[Momen et~al\mbox{.}(2020)]%
        {momen2020}
\bibfield{author}{\bibinfo{person}{Nurul Momen}, \bibinfo{person}{Sven Bock}, {and} \bibinfo{person}{Lothar Fritsch}.} \bibinfo{year}{2020}\natexlab{}.
\newblock \showarticletitle{Accept - Maybe - Decline: Introducing Partial Consent for the Permission-based Access Control Model of Android}. In \bibinfo{booktitle}{\emph{Proceedings of the 25th ACM Symposium on Access Control Models and Technologies}} (Barcelona, Spain) \emph{(\bibinfo{series}{SACMAT '20})}. \bibinfo{publisher}{Association for Computing Machinery}, \bibinfo{address}{New York, NY, USA}, \bibinfo{pages}{71–80}.
\newblock
\showISBNx{9781450375689}
\urldef\tempurl%
\url{https://doi.org/10.1145/3381991.3395603}
\showDOI{\tempurl}


\bibitem[Morrison et~al\mbox{.}(2014)]%
        {morrison2014improving}
\bibfield{author}{\bibinfo{person}{Alistair Morrison}, \bibinfo{person}{Donald McMillan}, {and} \bibinfo{person}{Matthew Chalmers}.} \bibinfo{year}{2014}\natexlab{}.
\newblock \showarticletitle{Improving consent in large scale mobile HCI through personalised representations of data}. In \bibinfo{booktitle}{\emph{Proceedings of the 8th Nordic Conference on Human-Computer Interaction: Fun, Fast, Foundational}} (Helsinki, Finland) \emph{(\bibinfo{series}{NordiCHI '14})}. \bibinfo{publisher}{Association for Computing Machinery}, \bibinfo{address}{New York, NY, USA}, \bibinfo{pages}{471–480}.
\newblock
\showISBNx{9781450325424}
\urldef\tempurl%
\url{https://doi.org/10.1145/2639189.2639239}
\showDOI{\tempurl}


\bibitem[Nair et~al\mbox{.}(2023a)]%
        {nair2023exploring}
\bibfield{author}{\bibinfo{person}{Vivek Nair}, \bibinfo{person}{Gonzalo~Munilla Garrido}, \bibinfo{person}{Dawn Song}, {and} \bibinfo{person}{James O'Brien}.} \bibinfo{year}{2023}\natexlab{a}.
\newblock \showarticletitle{Exploring the privacy risks of adversarial VR game design}.
\newblock \bibinfo{journal}{\emph{Proceedings on Privacy Enhancing Technologies}} (\bibinfo{year}{2023}).
\newblock


\bibitem[Nair et~al\mbox{.}(2023b)]%
        {nair2023uniqueID}
\bibfield{author}{\bibinfo{person}{Vivek Nair}, \bibinfo{person}{Wenbo Guo}, \bibinfo{person}{Justus Mattern}, \bibinfo{person}{Rui Wang}, \bibinfo{person}{James~F. O{\textquoteright}Brien}, \bibinfo{person}{Louis Rosenberg}, {and} \bibinfo{person}{Dawn Song}.} \bibinfo{year}{2023}\natexlab{b}.
\newblock \showarticletitle{Unique Identification of 50,000+ Virtual Reality Users from Head \& Hand Motion Data}. In \bibinfo{booktitle}{\emph{32nd USENIX Security Symposium (USENIX Security 23)}}. \bibinfo{publisher}{USENIX Association}, \bibinfo{address}{Anaheim, CA}, \bibinfo{pages}{895--910}.
\newblock
\showISBNx{978-1-939133-37-3}
\urldef\tempurl%
\url{https://www.usenix.org/conference/usenixsecurity23/presentation/nair-identification}
\showURL{%
\tempurl}


\bibitem[Obar and Oeldorf-Hirsch(2020)]%
        {Obar2020}
\bibfield{author}{\bibinfo{person}{Jonathan~A. Obar} {and} \bibinfo{person}{Anne Oeldorf-Hirsch}.} \bibinfo{year}{2020}\natexlab{}.
\newblock \showarticletitle{The biggest lie on the Internet: ignoring the privacy policies and terms of service policies of social networking services}.
\newblock \bibinfo{journal}{\emph{Information, Communication \& Society}} \bibinfo{volume}{23}, \bibinfo{number}{1} (\bibinfo{year}{2020}), \bibinfo{pages}{128--147}.
\newblock
\urldef\tempurl%
\url{https://doi.org/10.1080/1369118X.2018.1486870}
\showDOI{\tempurl}
\showeprint{https://doi.org/10.1080/1369118X.2018.1486870}


\bibitem[Paneva et~al\mbox{.}(2024)]%
        {paneva2024ieeepvc}
\bibfield{author}{\bibinfo{person}{Viktorija Paneva}, \bibinfo{person}{Marvin Strauss}, \bibinfo{person}{Verena Winterhalter}, \bibinfo{person}{Stefan Schneegass}, {and} \bibinfo{person}{Florian Alt}.} \bibinfo{year}{2024}\natexlab{}.
\newblock \showarticletitle{Privacy in the Metaverse:}.
\newblock \bibinfo{journal}{\emph{IEEE Pervasive Computing}} \bibinfo{volume}{23}, \bibinfo{number}{3} (\bibinfo{year}{2024}), \bibinfo{pages}{73--78}.
\newblock
\urldef\tempurl%
\url{https://doi.org/10.1109/MPRV.2024.3432953}
\showDOI{\tempurl}


\bibitem[Pfeuffer et~al\mbox{.}(2019)]%
        {pfeuffer2019chi}
\bibfield{author}{\bibinfo{person}{Ken Pfeuffer}, \bibinfo{person}{Matthias~J. Geiger}, \bibinfo{person}{Sarah Prange}, \bibinfo{person}{Lukas Mecke}, \bibinfo{person}{Daniel Buschek}, {and} \bibinfo{person}{Florian Alt}.} \bibinfo{year}{2019}\natexlab{}.
\newblock \showarticletitle{Behavioural Biometrics in VR: Identifying People from Body Motion and Relations in Virtual Reality}. In \bibinfo{booktitle}{\emph{Proceedings of the 2019 CHI Conference on Human Factors in Computing Systems}} (Glasgow, Scotland Uk) \emph{(\bibinfo{series}{CHI '19})}. \bibinfo{publisher}{Association for Computing Machinery}, \bibinfo{address}{New York, NY, USA}, \bibinfo{pages}{1–12}.
\newblock
\showISBNx{9781450359702}
\urldef\tempurl%
\url{https://doi.org/10.1145/3290605.3300340}
\showDOI{\tempurl}


\bibitem[Prange et~al\mbox{.}(2024)]%
        {prange2024soups}
\bibfield{author}{\bibinfo{person}{Sarah Prange}, \bibinfo{person}{Pascal Knierim}, \bibinfo{person}{Gabriel Knoll}, \bibinfo{person}{Felix Dietz}, \bibinfo{person}{Alexander~De Luca}, {and} \bibinfo{person}{Florian Alt}.} \bibinfo{year}{2024}\natexlab{}.
\newblock \showarticletitle{I do (not) need that Feature! – Understanding Users’ Awareness and Control of Privacy Permissions on Android Smartphones}. In \bibinfo{booktitle}{\emph{Twentieth Symposium on Usable Privacy and Security (SOUPS 2024)}}. \bibinfo{publisher}{USENIX Association}, \bibinfo{address}{Philadelphia, PA}.
\newblock
\urldef\tempurl%
\url{https://www.usenix.net/conference/soups2024/presentation/prange}
\showURL{%
\tempurl}


\bibitem[Prange et~al\mbox{.}(2021)]%
        {prange2021priview}
\bibfield{author}{\bibinfo{person}{Sarah Prange}, \bibinfo{person}{Ahmed Shams}, \bibinfo{person}{Robin Piening}, \bibinfo{person}{Yomna Abdelrahman}, {and} \bibinfo{person}{Florian Alt}.} \bibinfo{year}{2021}\natexlab{}.
\newblock \showarticletitle{PriView– Exploring Visualisations to Support Users’ Privacy Awareness}. In \bibinfo{booktitle}{\emph{Proceedings of the 2021 CHI Conference on Human Factors in Computing Systems}} (Yokohama, Japan) \emph{(\bibinfo{series}{CHI '21})}. \bibinfo{publisher}{Association for Computing Machinery}, \bibinfo{address}{New York, NY, USA}, Article \bibinfo{articleno}{69}, \bibinfo{numpages}{18}~pages.
\newblock
\showISBNx{9781450380966}
\urldef\tempurl%
\url{https://doi.org/10.1145/3411764.3445067}
\showDOI{\tempurl}


\bibitem[Rajaram et~al\mbox{.}(2023a)]%
        {Rajaram2023}
\bibfield{author}{\bibinfo{person}{Shwetha Rajaram}, \bibinfo{person}{Chen Chen}, \bibinfo{person}{Franziska Roesner}, {and} \bibinfo{person}{Michael Nebeling}.} \bibinfo{year}{2023}\natexlab{a}.
\newblock \showarticletitle{Eliciting Security \& Privacy-Informed Sharing Techniques for Multi-User Augmented Reality}. In \bibinfo{booktitle}{\emph{Proceedings of the 2023 CHI Conference on Human Factors in Computing Systems}} (Hamburg, Germany) \emph{(\bibinfo{series}{CHI '23})}. \bibinfo{publisher}{Association for Computing Machinery}, \bibinfo{address}{New York, NY, USA}, Article \bibinfo{articleno}{98}, \bibinfo{numpages}{17}~pages.
\newblock
\showISBNx{9781450394215}
\urldef\tempurl%
\url{https://doi.org/10.1145/3544548.3581089}
\showDOI{\tempurl}


\bibitem[Rajaram et~al\mbox{.}(2023b)]%
        {rajaram2023reframe}
\bibfield{author}{\bibinfo{person}{Shwetha Rajaram}, \bibinfo{person}{Franziska Roesner}, {and} \bibinfo{person}{Michael Nebeling}.} \bibinfo{year}{2023}\natexlab{b}.
\newblock \showarticletitle{Reframe: An Augmented Reality Storyboarding Tool for Character-Driven Analysis of Security \& Privacy Concerns}. In \bibinfo{booktitle}{\emph{Proceedings of the 36th Annual ACM Symposium on User Interface Software and Technology}} (San Francisco, CA, USA) \emph{(\bibinfo{series}{UIST '23})}. \bibinfo{publisher}{Association for Computing Machinery}, \bibinfo{address}{New York, NY, USA}, Article \bibinfo{articleno}{117}, \bibinfo{numpages}{15}~pages.
\newblock
\showISBNx{9798400701320}
\urldef\tempurl%
\url{https://doi.org/10.1145/3586183.3606750}
\showDOI{\tempurl}


\bibitem[Rajivan and Camp(2016)]%
        {rajivan2016influence}
\bibfield{author}{\bibinfo{person}{Prashanth Rajivan} {and} \bibinfo{person}{Jean Camp}.} \bibinfo{year}{2016}\natexlab{}.
\newblock \showarticletitle{Influence of Privacy Attitude and Privacy Cue Framing on Android App {Choices}}. In \bibinfo{booktitle}{\emph{Twelfth Symposium on Usable Privacy and Security (SOUPS 2016)}}. \bibinfo{publisher}{USENIX Association}, \bibinfo{address}{Denver, CO}.
\newblock
\urldef\tempurl%
\url{https://www.usenix.org/conference/soups2016/workshop-program/wpi/presentation/rajivan}
\showURL{%
\tempurl}


\bibitem[Reilly et~al\mbox{.}(2014)]%
        {Reilly2014}
\bibfield{author}{\bibinfo{person}{Derek Reilly}, \bibinfo{person}{Mohamad Salimian}, \bibinfo{person}{Bonnie MacKay}, \bibinfo{person}{Niels Mathiasen}, \bibinfo{person}{W.~Keith Edwards}, {and} \bibinfo{person}{Juliano Franz}.} \bibinfo{year}{2014}\natexlab{}.
\newblock \showarticletitle{SecSpace: prototyping usable privacy and security for mixed reality collaborative environments}. In \bibinfo{booktitle}{\emph{Proceedings of the 2014 ACM SIGCHI Symposium on Engineering Interactive Computing Systems}} (Rome, Italy) \emph{(\bibinfo{series}{EICS '14})}. \bibinfo{publisher}{Association for Computing Machinery}, \bibinfo{address}{New York, NY, USA}, \bibinfo{pages}{273–282}.
\newblock
\showISBNx{9781450327251}
\urldef\tempurl%
\url{https://doi.org/10.1145/2607023.2607039}
\showDOI{\tempurl}


\bibitem[Rothchild(2017)]%
        {rothchild2017against}
\bibfield{author}{\bibinfo{person}{John~A Rothchild}.} \bibinfo{year}{2017}\natexlab{}.
\newblock \showarticletitle{Against notice and choice: The manifest failure of the proceduralist paradigm to protect privacy online (or anywhere else)}.
\newblock \bibinfo{journal}{\emph{Clev. St. L. Rev.}}  \bibinfo{volume}{66} (\bibinfo{year}{2017}), \bibinfo{pages}{559}.
\newblock


\bibitem[Ruth et~al\mbox{.}(2019)]%
        {Ruth2019}
\bibfield{author}{\bibinfo{person}{Kimberly Ruth}, \bibinfo{person}{Tadayoshi Kohno}, {and} \bibinfo{person}{Franziska Roesner}.} \bibinfo{year}{2019}\natexlab{}.
\newblock \showarticletitle{Secure {Multi-User} Content Sharing for Augmented Reality Applications}. In \bibinfo{booktitle}{\emph{28th USENIX Security Symposium (USENIX Security 19)}}. \bibinfo{publisher}{USENIX Association}, \bibinfo{address}{Santa Clara, CA}, \bibinfo{pages}{141--158}.
\newblock
\showISBNx{978-1-939133-06-9}
\urldef\tempurl%
\url{https://www.usenix.org/conference/usenixsecurity19/presentation/ruth}
\showURL{%
\tempurl}


\bibitem[Sloan and Warner(2014)]%
        {sloan2014beyond}
\bibfield{author}{\bibinfo{person}{Robert~H Sloan} {and} \bibinfo{person}{Richard Warner}.} \bibinfo{year}{2014}\natexlab{}.
\newblock \showarticletitle{Beyond notice and choice: Privacy, norms, and consent}.
\newblock \bibinfo{journal}{\emph{J. High Tech. L.}}  \bibinfo{volume}{14} (\bibinfo{year}{2014}), \bibinfo{pages}{370}.
\newblock


\bibitem[Steil et~al\mbox{.}(2019)]%
        {steil2019etra}
\bibfield{author}{\bibinfo{person}{Julian Steil}, \bibinfo{person}{Inken Hagestedt}, \bibinfo{person}{Michael~Xuelin Huang}, {and} \bibinfo{person}{Andreas Bulling}.} \bibinfo{year}{2019}\natexlab{}.
\newblock \showarticletitle{Privacy-Aware Eye Tracking Using Differential Privacy}. In \bibinfo{booktitle}{\emph{Proceedings of the 11th ACM Symposium on Eye Tracking Research \& Applications}} (Denver, Colorado) \emph{(\bibinfo{series}{ETRA '19})}. \bibinfo{publisher}{Association for Computing Machinery}, \bibinfo{address}{New York, NY, USA}, Article \bibinfo{articleno}{27}, \bibinfo{numpages}{9}~pages.
\newblock
\showISBNx{9781450367097}
\urldef\tempurl%
\url{https://doi.org/10.1145/3314111.3319915}
\showDOI{\tempurl}


\bibitem[Tabbaa et~al\mbox{.}(2022)]%
        {Tabbaa2022}
\bibfield{author}{\bibinfo{person}{Luma Tabbaa}, \bibinfo{person}{Ryan Searle}, \bibinfo{person}{Saber~Mirzaee Bafti}, \bibinfo{person}{Md~Moinul Hossain}, \bibinfo{person}{Jittrapol Intarasisrisawat}, \bibinfo{person}{Maxine Glancy}, {and} \bibinfo{person}{Chee~Siang Ang}.} \bibinfo{year}{2022}\natexlab{}.
\newblock \showarticletitle{VREED: Virtual Reality Emotion Recognition Dataset Using Eye Tracking \& Physiological Measures}.
\newblock \bibinfo{journal}{\emph{Proc. ACM Interact. Mob. Wearable Ubiquitous Technol.}} \bibinfo{volume}{5}, \bibinfo{number}{4}, Article \bibinfo{articleno}{178} (\bibinfo{date}{Dec.} \bibinfo{year}{2022}), \bibinfo{numpages}{20}~pages.
\newblock
\urldef\tempurl%
\url{https://doi.org/10.1145/3495002}
\showDOI{\tempurl}


\bibitem[Tahaei et~al\mbox{.}(2023)]%
        {tahaei2023stuck}
\bibfield{author}{\bibinfo{person}{Mohammad Tahaei}, \bibinfo{person}{Ruba Abu-Salma}, {and} \bibinfo{person}{Awais Rashid}.} \bibinfo{year}{2023}\natexlab{}.
\newblock \showarticletitle{Stuck in the permissions with you: Developer \& end-user perspectives on app permissions \& their privacy ramifications}. In \bibinfo{booktitle}{\emph{Proceedings of the 2023 CHI Conference on Human Factors in Computing Systems}}. \bibinfo{pages}{1--24}.
\newblock


\bibitem[Trimananda et~al\mbox{.}(2022)]%
        {trimananda2022ovrseen}
\bibfield{author}{\bibinfo{person}{Rahmadi Trimananda}, \bibinfo{person}{Hieu Le}, \bibinfo{person}{Hao Cui}, \bibinfo{person}{Janice~Tran Ho}, \bibinfo{person}{Anastasia Shuba}, {and} \bibinfo{person}{Athina Markopoulou}.} \bibinfo{year}{2022}\natexlab{}.
\newblock \showarticletitle{{OVRseen}: Auditing Network Traffic and Privacy Policies in Oculus {VR}}. In \bibinfo{booktitle}{\emph{31st USENIX Security Symposium (USENIX Security 22)}}. \bibinfo{publisher}{USENIX Association}, \bibinfo{address}{Boston, MA}, \bibinfo{pages}{3789--3806}.
\newblock
\showISBNx{978-1-939133-31-1}
\urldef\tempurl%
\url{https://www.usenix.org/conference/usenixsecurity22/presentation/trimananda}
\showURL{%
\tempurl}


\bibitem[Volk et~al\mbox{.}(2022)]%
        {volk2022pricheck}
\bibfield{author}{\bibinfo{person}{Vera Volk}, \bibinfo{person}{Sarah Prange}, {and} \bibinfo{person}{Florian Alt}.} \bibinfo{year}{2022}\natexlab{}.
\newblock \showarticletitle{PriCheck– An Online Privacy Assistant for Smart Device Purchases}. In \bibinfo{booktitle}{\emph{Extended Abstracts of the 2022 CHI Conference on Human Factors in Computing Systems}} (New Orleans, LA, USA) \emph{(\bibinfo{series}{CHI EA '22})}. \bibinfo{publisher}{Association for Computing Machinery}, \bibinfo{address}{New York, NY, USA}, Article \bibinfo{articleno}{275}, \bibinfo{numpages}{5}~pages.
\newblock
\showISBNx{9781450391566}
\urldef\tempurl%
\url{https://doi.org/10.1145/3491101.3519827}
\showDOI{\tempurl}


\bibitem[Zhang et~al\mbox{.}(2018)]%
        {Zhang2018}
\bibfield{author}{\bibinfo{person}{Yongtuo Zhang}, \bibinfo{person}{Wen Hu}, \bibinfo{person}{Weitao Xu}, \bibinfo{person}{Chun~Tung Chou}, {and} \bibinfo{person}{Jiankun Hu}.} \bibinfo{year}{2018}\natexlab{}.
\newblock \showarticletitle{Continuous Authentication Using Eye Movement Response of Implicit Visual Stimuli}.
\newblock \bibinfo{journal}{\emph{Proc. ACM Interact. Mob. Wearable Ubiquitous Technol.}} \bibinfo{volume}{1}, \bibinfo{number}{4}, Article \bibinfo{articleno}{177} (\bibinfo{date}{Jan.} \bibinfo{year}{2018}), \bibinfo{numpages}{22}~pages.
\newblock
\urldef\tempurl%
\url{https://doi.org/10.1145/3161410}
\showDOI{\tempurl}


\bibitem[Zhu et~al\mbox{.}(2020)]%
        {Zhu2020}
\bibfield{author}{\bibinfo{person}{Huadi Zhu}, \bibinfo{person}{Wenqiang Jin}, \bibinfo{person}{Mingyan Xiao}, \bibinfo{person}{Srinivasan Murali}, {and} \bibinfo{person}{Ming Li}.} \bibinfo{year}{2020}\natexlab{}.
\newblock \showarticletitle{BlinKey: A Two-Factor User Authentication Method for Virtual Reality Devices}.
\newblock \bibinfo{journal}{\emph{Proc. ACM Interact. Mob. Wearable Ubiquitous Technol.}} \bibinfo{volume}{4}, \bibinfo{number}{4}, Article \bibinfo{articleno}{164} (\bibinfo{date}{Dec.} \bibinfo{year}{2020}), \bibinfo{numpages}{29}~pages.
\newblock
\urldef\tempurl%
\url{https://doi.org/10.1145/3432217}
\showDOI{\tempurl}


\end{thebibliography}

\appendix

\section{Semi-structured Interview}
\label{semi-structure_interview}
\begin{enumerate}
    \item Summarise your data collection awareness and control design in your own words.
Was it difficult to come up with the ideas? Why?
    \item How would your design change across different types of games (e.g., FPS, platformer, strategy, casual etc.) and across different XR applications (e.g., entertainment, education, training)?
    \item Do you have concerns about the potential for uniquely identifying users from gaming data, both as a player and a game designer?
    \item What would be some good strategies to incorporate user privacy from the onset? What do you think are the main challenges to privacy-aware design?
\end{enumerate}

\end{document}